\documentclass[manuscript]{acmart}

\usepackage{booktabs}
\usepackage{tabularx}
\usepackage{array}

\newcolumntype{Y}{>{\raggedright\arraybackslash}X}

\AtBeginDocument{%
  }

\setcopyright{acmlicensed}
\copyrightyear{2018}
\acmYear{2018}
\acmDOI{XXXXXXX.XXXXXXX}
\acmConference[Conference acronym 'XX]{Make sure to enter the correct
  conference title from your rights confirmation email}{June 03--05,
  2018}{Woodstock, NY}
\acmISBN{978-1-4503-XXXX-X/2018/06}

\begin{document}

\title{The Paradox of Agency-Preserving AI: Negotiating Learner Agency in Layered Human–AI Interactions}


\author{Yueling Fan}
\orcid{0009-0004-0737-9196}
\affiliation{%
  \institution{KTH Royal Institute of Technology}
  \city{Stockholm}
  \country{Sweden}
}

\author{Jacqueline Wong}
\orcid{0000-0002-5387-7696}
\affiliation{%
  \institution{Utrecht University}
  \city{Utrecht}
  \country{The Netherlands}
}
\affiliation{%
  \institution{Digital Futures}
  \city{Stockholm}
  \country{Sweden}
}
\email{l.y.j.wong@uu.nl}

\author{Olov Engwall}
\orcid{0000-0003-4532-014X}
\affiliation{%
  \institution{KTH Royal Institute of Technology}
  \city{Stockholm}
  \country{Sweden}
}
\affiliation{%
  \institution{Digital Futures}
  \city{Stockholm}
  \country{Sweden}
}

\author{Olga Viberg}
\orcid{0000-0002-8543-3774}
\correspondingauthor
\email{oviberg@kth.se}
\affiliation{%
  \institution{KTH Royal Institute of Technology}
  \city{Stockholm}
  \country{Sweden}
}
\affiliation{%
  \institution{Digital Futures}
  \city{Stockholm}
  \country{Sweden}
}

\renewcommand{\shortauthors}{Fan et al.}


\begin{abstract}
Generative AI is increasingly used by learners in cognitively demanding tasks, but its capacity to provide immediate solutions raises concerns about decreased learner agency. Meta-level AI systems that provide reflective and metacognitive scaffolding have emerged as a promising approach to supporting learners' self-regulation. However, these systems introduce a fundamental paradox: AI intended to preserve learner agency can itself constrain that agency by regulating \textit{when, how,} and \textit{why }learners should reflect. 
To investigate this challenge, we conducted a co-design workshop with 18 students engaged in programming and academic writing tasks using ChatGPT extended with \textit{ChatWise}, a meta-level AI system providing self-regulated learning-aligned feedback. Drawing on 185 prompt–feedback entries, 174 written reflections, and participants' design artifacts (17 storyboards and 5 group design proposals), we identified four interconnected design tensions: assistance vs reflection, immediate intervention vs space for self-regulation, visibility vs attention competition, and system adaptation vs user control. 
The study contributes a multidimensional account of learner agency that includes task, cognitive, and interaction agency, along with empirically grounded interaction-design mechanisms to negotiate these tensions. 
  
\end{abstract}


\received{20 February 2007}
\received[revised]{12 March 2009}
\received[accepted]{5 June 2009}

\maketitle

\noindent Word count: 10520 (excluding references)

\section{Introduction}
Generative AI (GenAI) offers learners immediate explanations, solutions, and feedback on cognitively demanding tasks \cite{lee2024harnessing,klein2026answer,isa2026codegeneration}, which creates opportunities for personalized learning and timely assistance but may also encourage learners to offload cognitive activities central to learning, including problem formulation, reasoning, problem-solving, and critical evaluation \cite{jelson2026empirical,isa2026codegeneration}.
The fluent responses and frictionless interactions offered by commercial conversational GenAI systems may nudge learners towards immediate task completion at the expense of sustained cognitive engagement \cite{klein2026answer,li2026supporting}. These issues raise fundamental questions about learner agency: who decides which cognitive activities should be delegated to AI, which should remain learner-led, and when AI assistance is appropriate \cite{borchers2026whodecides,song2026goodfences}. Rather than treating learner agency simply as the ability to interact with or control an AI system, recent human-computer interaction (HCI) research conceptualizes it as \textit{relational, contextual}, and \textit{dynamically }configured through the distribution of decision authority between humans and AI systems \cite{zhang2025collaboration,bennett2023agency,borchers2026whodecides}. Meaningful learner \textit{agency} extends beyond access to choices or interface controls to learners' ability to exercise consequential decision-making and influence how their interactions with AI unfold \cite{pauzi2026evaluating,borchers2026whodecides}. This authority may concern different aspects of the learning experience, including task progression, cognitive participation, and the management of AI interactions \cite{chucarroll1997tracking,taylor2026contextmaxxing}. 

To address concerns about excessive cognitive delegation and preserve learner agency, researchers have explored AI-based scaffolding designed to foster self-regulated learning (SRL), including self-monitoring and reflection \cite{pintrich2004conceptual}. Existing approaches include modifying conversational AI behavior to encourage critical evaluation \cite{li2026supporting}, introducing reflective scaffolds grounded in pedagogical theories such as the Zone of Proximal Development \cite{vygotsky1978mind} and Self-Determination Theory \cite{ryan2000self,  sharma2026hybrid}, and integrating SRL-aligned interventions that prompt learners to articulate goals, monitor understanding, and evaluate learning strategies \cite{klein2026answer, wang2025srlmentor}. While these approaches may foster SRL and strengthen learners' sense of agency, they can also impose additional cognitive demands, undermine motivation, or distrupt task progression \cite{tian2026critical,klein2026answer,bucinca2021trust}. Moreover, overly prescriptive or poorly contextualised scaffolding risks shifting regulatory control from the learner to the AI system \cite{molenaar2022hybrid}. This creates a paradox: AI systems designed to preserve learner agency may inadvertently constrain it by determining \textit{when, what,} and \textit{how }learners should reflect \cite{bucinca2021trust,tsai2020empowering,salah2026regulation}.

This paradox becomes particularly salient in layered AI interactions, where learners engage with a task-oriented primary AI alongside a secondary, meta-level AI intended to support reflection and regulate engagement with the primary system. Whereas primary AI supports task accomplishment, meta-level AI scaffolds how learners approach, evaluate, and regulate that assistance. Recent research demonstrates the potential of external metacognitive agents to promote critical evaluation and more deliberate engagement with AI-generated outputs \cite{gmeiner2025metacognitive,park2026llmbox}. However, introducing this additional layer creates a distinct challenge for the distribution of decision authority: learners must regulate not only which cognitive activities to delegate to the primary AI, but also \textit{whether, when}, and \textit{how} to engage with the meta-level AI's guidance. As these layers may introduce competing demands on learners' attention, reasoning, and task progression, the central design challenge extends beyond providing effective reflective scaffolding to ensuring learners retain meaningful control over how such scaffolding shapes their learning processes. 


Prior research suggests that the effectiveness of reflective scaffolding depends on the timing, presentation, and adaptation of interventions to learners' needs \cite{tian2026critical,dhillon2024shaping,chen2026passive}. Feedback delivered after a complete AI-generated solution may arrive too late to influence reasoning, whereas interventions introduced prematurely or too frequently may disrupt task progression \cite{klein2026answer,zhang2026its}. Similarly, highly visible or information-dense feedback may compete for attention with the primary task \cite{li2026supporting,cheng2025noticeability}, while adaptive interventions raise questions about who determines what support is appropriate and how it should evolve \cite{song2026goodfences}. From an HCI perspective, these challenges highlight the need to examine not only the pedagogical content of reflective feedback but also \textit{how} interaction mechanisms can enable learners to shape its timing, visibility, intensity, and adaptation. In layered AI environments, such mechanisms become particularly important because they influence the extent to which learners can actively regulate their engagement with meta-level AI support, rather than merely respond to interventions initiated by the system. 

Building on earlier work \cite{viberg2026scalable,fan2026learneragency}, we investigate \textit{ChatWise}, a lightweight browser extension that embeds SRL-aligned metacognitive feedback \cite{pintrich2004conceptual} into learners' interactions with commercial conversational AI systems without modifying the primary AI's responses. Previous evaluations provided initial evidence that \textit{ChatWise} could encourage prompt refinement, strategic awareness, and reflective engagement with AI-supported learning \cite{viberg2026scalable}. A subsequent participatory design study highlighted how feedback timing, interaction mode, and intervention strategy shaped learners' perceptions of reflective support \cite{fan2026learneragency}. In particular, learners were often less motivated to engage with metacognitive feedback after receiving ready-made solutions from the primary AI system, suggesting that the value of reflection depends on its relationship to ongoing task assistance. However, these findings leave open a broader HCI question: \textit{How do learners experience and negotiate decision authority across task-oriented and meta-level AI, and what interaction mechanisms could enable them to retain meaningful control over reflective support?}





This study makes \textbf{three contributions to HCI research on AI-mediated learning.} \textit{First}, we provide an empirical account of four interconnected agency-related design tensions that emerge when task-oriented generative AI operates alongside reflective meta-level AI, illustrating how interventions intended to support learning can simultaneously constrain learners' decision-making.\textit{ Second}, we advance a multidimensional account of learner agency in layered AI interactions by distinguishing task, cognitive-, and interaction agency, and demonstrating how these forms of decision authority can become interdependent or conflicting. \textit{Third}, drawing on participants' experiences and co-design proposals, we synthesize five empirically grounded design criteria for agency-preserving meta-level AI, supported by concrete interaction mechanisms for contextualizing, timing, presenting, revisiting, and personalizing reflective guidance. Together, these contributions reposition agency-preserving AI design as a challenge to negotiate decision authority across AI systems, rather than simply increasing the amount or intensity of reflective scaffolding.

\section{Background}

\subsection{Cognitive offloading, reflection, and scaffolding with GenAI} \label{section_2.2}


Given the convenience of immediate and often complete solutions offered by commercial LLMs (e.g., Claude or ChatGPT), learners favoring convenient shortcuts may (in)advertently delegate cognitive work embedded in learning tasks \cite{klein2026answer}. For example, in their study, Jelson and colleagues showed that many students tend to delegate most of the writing process to AI, ranging from topic exploration and planning to text generation and revision, with some students delegating the entire essay-writing task \cite{jelson2026empirical}. However, even students who retained some involvement in the writing process outsource some critical thinking processes such as ideation and opinion formation \cite{jelson2026empirical}. Similarly, scholars examined how students used LLMs to delegate programming-related work, including code generation, debugging, and conceptual explanation \cite{isa2026codegeneration}. They found that behaviors that require greater effort from the learner, such as providing detailed specifications for code generation, were positively associated with better learning performance, while lower-effort behaviors, such as pasting raw error messages for the model to resolve, were negatively associated with learning outcomes \cite{isa2026codegeneration}. These patterns raise concerns about cognitive offloading \cite{risko2016cognitive} and 'metacognitive laziness' \cite{fan2025metacognitive}. When learners can easily delegate cognitive effort to AI to improve task efficiency, they can miss opportunities to participate in SRL—planning and selecting suitable learning strategies, monitoring their understanding, and evaluating and reflecting on their learning process \cite{hayes1986writing}.

Existing research has explored different approaches that introduce deliberate friction in AI-mediated learning to support reflective engagement. Li and colleagues for example, redesigned a chatbot to reconfigure the behavior of the model, such as expressing uncertainty and reducing verbosity and found  that reduced verbosity can improve the detection of logically flawed responses by learners, while consistently expressing uncertainty can reduce the adoption of incorrect chatbot responses \cite{li2026supporting}. Other work has drawn on learning theories to guide the design of scaffolds for learner engagement. For instance, Sharma et al. developed a hybrid dialogue system combining a rule-based structure based on SRL with LLM-generated, context-sensitive follow-up questions to support learner reflection \cite{sharma2026hybrid}. Scholars have also shown that in interactions with a social robot, students were more likely to detect flawed arguments when the robot expressed uncertainty, but they nevertheless had an overtrust in the robot's solution; uncertainty may hence help, but it cannot be the only solution \cite{gonzalez2025sense}. 

Previous work on reflective scaffolding has highlighted challenges related to disengagement, including increased mental demands due to information overload and processing difficulties \cite{tian2026critical}, distrust arising from misjudgments of learners’ intentions and frequent uncertain responses \cite{sharma2026hybrid,li2026supporting}, and reduced confidence resulting from excessive criticism \cite{tian2026critical}. These challenges highlight the risks of designing scaffolding with overly strict control or becoming coercive or authoritative \cite{klein2026answer}. In this study, we study an external meta-level metacognitive scaffolding while avoiding interference with learners' use of task-oriented primary AI. 
We aim to explore design tensions and corresponding interaction mechanisms for scaffolding that support learners in regulating cognitive delegation, enabling them to judge and decide which cognitive activities to delegate, which to retain, and how to delegate in ways that are conducive to their learning.




\subsection{Learner agency and delegation in AI-mediated learning}


When GenAI systems are designed to take on increasingly proactive roles in interaction, they may initiate actions, provide recommendations, and adapt their behavior in response to evolving contexts. Agency therefore is configured through how human and AI roles are arranged (i.e., \textit{relational}) and the opportunities each has to act, respond, redirect interaction, and influence outcomes (i.e., \textit{contextual}). The distribution of agency may vary across interaction contexts and change as interaction unfolds (i.e., \textit{dynamic})
)\cite{yun2026chatbot,wu2026afterinterface}. In the AI-mediated learning contexts, scholars argue that learner agency should be explained in terms of whether learners can make consequential choices and meaningfully influence subsequent system behavior or learning pathways \cite{pauzi2026evaluating}. This means that learners have the option to reject a system recommendation, and the system would adjust subsequent tasks or feedback based on the learners' decisions.

Furthermore, scholars have emphasized the need to distinguish between \textit{the task initiative}, which concerns leading the development of a collaborative task, and \textit{the dialogue initiative}, which concerns leading the current focus of the discourse \cite{chucarroll1997tracking}. In their analysis, they have shown that a participant can take the conversational lead without taking the lead in directing how the task should be accomplished \cite{chucarroll1997tracking}. Scholars also suggest that the learner initiative or decision-making in mixed-initiative situations may have different purposes, such as reasoning and reflection, task execution, or dialogue management \cite{song2026goodfences,isa2026codegeneration,sharma2026hybrid}. They further argue that tracking the distribution of these two forms of initiative can help systems adapt their responses and support coordination between users and the system \cite{chucarroll1997tracking}. Therefore, distinguishing between these initiatives and tracking the distribution is crucial, yet remain under-explored. 

As GenAI closely engages with human reasoning and intentionality, Taylor and Krishna propose \textit{cognitive agency} in terms of people’s control, efficacy, and mastery of thinking and acting with AI. It emphasizes the value of metacognitive monitoring of user behavior in their own thinking, exercise of judgment, and development of understanding through collaboration with AI \cite{taylor2026contextmaxxing}. As learner agency is \textit{multidimensional} and \textit{interactional}, shaped through systems' affordances, decision policies, feedback structures, and orchestration behaviors \cite{pauzi2026evaluating}, the distinction between \textit{task},\textit{ cognitive}, and \textit{interaction }dimensions is crucial for increasing our deeper understanding of students' needs for agency-preserving AI learner support.

However, examining learner agency in practice is particularly challenging as observable interaction alone may not fully reveal \textit{how} learners intend to allocate work and decision authority. Previous studies on large language models'(LLM) delegation demonstrate that learners can face difficulties both in selecting appropriate delegation strategies and in carrying out their intended boundaries \cite{song2026goodfences}. The Agency Allocation Framework (AAF) \cite{borchers2026whodecides} provides a structured approach to operationalizing learner agency. It directs attention to specific decision points by specifying what decisions are being examined, who participates and with what authority, how the available choices are structured, and what evidence supports claims about agency. Building on this perspective, the study examines the learning-related decisions students seek to retain, delegate, or contest, together with the interaction mechanisms that shape how those decisions are enacted in their interactions with conversational AI and meta-level scaffolding during the programming and writing tasks (Section \ref{individual_tasks}).

\subsection{Agency as an interaction-design problem: timing, attention, and control} \label{2.3}

 Taking the HCI's lens of agency, reflective scaffolding that preserves agency would need to consider \textit{how} the interaction and system design choices are configured to provide timely, controllable, and adaptive support. In this section, we focus on three key aspects: (1) temporal organization, (2) attention and visibility, and (3) adaptation and user control.
 

\textbf{Temporal organization}.
While GenAI is able to offer immediate metacognitive feedback, research has highlighted the need to preserve reflective pauses that allow learners to engage in retrospective self-reflection and evaluation. For example, Zhao and colleagues developed a self-regulated gamified VR system that provides immediate visual feedback for quick error correction during English learning tasks. After each learning unit, learners were guided to assess their expected performance, actual learning time, and learning strategies before receiving system-generated overall diagnostic feedback \cite{zhao2024gamifiedvr}. Students who learned with this system showed better English performance and self-reported self-regulation than those using gamified VR without the self-regulation support. In another study, \textit{SRLMentor}, an SRL-supportive GenAI system, similarly provided immediate diagnostic feedback to help learners identify knowledge gaps, while also sending summarized scheduled feedback through periodic emails to provide learners with more space for independent reflection \cite{wang2025srlmentor}. Furthermore, Zhang et al. found that this type of immediate feedback paired with delayed feedback timing shows the largest observed effect compared to either the immediate- or delayed-only feedback in their meta-analysis \cite{zhang2026its}. They argue that such designs may support both reflection-in-action, where learners monitor and adjust their approaches during the ongoing task, and reflection-on-action, where they revisit and evaluate their learning processes after completing it \cite{schon1983reflective}.

\textbf{Attention and visibility. }
In GenAI-mediated learning settings, reflective scaffolding faces an attentional trade-off: guidance should be salient and detailed enough to support reflection \cite{zhang2025friction}, but overly detailed or prominent AI explanations can increase cognitive load and lead to negative impact \cite{li2026supporting,anik2026dataset,darvishi2024agency,cheng2025noticeability}. Building on the principle of calm technology, when information can move between the center and periphery of attention \cite{weiser1996calm}, prior HCI work has explored ambient learning displays \cite{borner2014lead} and progressive disclosure \cite{anik2026dataset,liu2026tasklens} approaches to regulate the attentional demands of supportive information. In layered AI interaction, this becomes especially challenging as the noticeability of secondary information also needs to be balanced against its potential to demand focal attention \cite{cheng2025noticeability}.
In the present study, therefore, the important issue is that a meta-AI has to be sufficiently visible to influence behavior without competing with the primary AI for focal attention.

\textbf{Adaptation and user control.}
Learners differ in how they engage with AI support. Research on AI-mediated writing indicates that learners' cognitive delegation patterns vary with factors such as learning self-efficacy and the perceived usefulness of AI \cite{jelson2026empirical}. Learners may also respond differently to interactive AI-supported environments, depending on cognitive characteristics such as cognitive flexibility, i.e., the ability and willingness to consider alternatives and adapt their responses to changing situations \cite{wu2026interactive}. These differences suggest that effective learner agency-preserving GenAI scaffolding needs to adapt and personalize support to different learner characteristics rather than provide uniform guidance. However, adaptation also requires the system to make judgments about what kind of support is appropriate for a learner. This can raise problems because the cognitive work learners intend to retain or delegate can differ depending on their learning goals, prior experiences, and situational concerns \cite{song2026goodfences}. Adaptive support therefore needs to remain open for learners to contest, correct, and steer how support is personalized.

Together, these three dimensions highlight that preserving learner agency is not only a matter of what reflective support is provided, but also when, how, and under whose control it is delivered. These design considerations become particularly important when reflective scaffolding is introduced through a separate AI system that monitors and intervenes in learners' interactions with a primary AI. While such meta-level support may encourage reflection and more deliberate engagement, it also introduces a potential tension: an AI system designed to preserve learner agency may itself constrain that agency through its decisions about when and how to intervene.

\subsection{Meta-Level AI for Regulating Human–AI Interaction}




One emerging approach to providing such reflective scaffolding involves meta-level AI systems that operate alongside a primary AI, supporting users in monitoring and regulating their interactions rather than directly assisting with task completion. Recent work has explored the potential of these systems to manage cognitive offloading and support human agency in cognitively demanding tasks. For example, Gmeiner et al. have introduced metacognitive support agents that operate alongside the primary AI system, Autodesk Fusion 360’s Generative Design tool \cite{gmeiner2025metacognitive}. These agents aim to help users regulate the cognitive processes involved in formulating design intent, exploring the design problem, and evaluating AI-generated outcomes. Participants receiving such support generally produced more feasible designs, with the interventions prompting them to consider previously overlooked design factors and revise inputs to the generative design system. Park et al. instead have focused on users’ verification of AI outputs \cite{park2026llmbox}. Their Thinking-box presents sentence-level reflective checkpoints derived from patterns of distorted information. The intervention has shown positive effects, such as encouraging more deliberate scrutiny of LLM responses and reducing missed instances of distorted information relative to the baseline condition.

In line with this perspective, \textit{ChatWise} focuses on regulating how learners engage with commercial conversational AI rather than modifying the primary AI itself \cite{viberg2026scalable}. It combines learning analytics with SRL-aligned conversational scaffolding to classify learners’ prompts and provide adaptive metacognitive feedback in real time. In a within-subject field study, learners using \textit{ChatWise} produced higher-quality prompts more than twice as often and showed greater prompt refinement, strategic awareness, and reflective engagement with AI-supported learning. However, the impact of introducing a meta-level AI into an ongoing learner-AI dialogue on learners' conversational experience and sense of agency remains under-explored. From the perspective of \textit{dialogic intelligence}, such interventions may shape how dialogue is opened, widened, deepened, sustained, and applied \cite{wegerif2026dialogic}. This motivates us to explore how learners negotiate agency when engaging with a primary AI oriented toward task completion alongside meta-level feedback intended to prompt metacognitive reflection.

\subsection{Research Gap}
Taken together, prior work demonstrates the potential of reflective and meta-level AI interventions to support learners in regulating their engagement with generative AI. However, we still know relatively little about how introducing a meta-level AI alongside a task-oriented primary AI reshapes learner agency. In particular, existing work has not sufficiently examined how learners negotiate decision authority across these two AI layers, or how the interaction design of meta-level guidance—including its timing, visibility, and adaptation—can support reflection without itself becoming overly prescriptive, interruptive, or autonomous. 

To address this gap, this study asks:

\textbf{RQ1:} How do learners perceive and negotiate decision authority over task progression, cognitive engagement, and AI interaction when engaging with task-oriented primary and meta-level secondary AI systems?

\textbf{RQ2:} What interaction-design mechanisms that can support meta-level AI in promoting reflection and self-regulation while preserving learner agency emerge?



\section{Method}
We conducted a co-design workshop to investigate how learners negotiate decision authority when meta-level AI guidance is layered onto their ongoing interaction with a task-oriented primary AI. The study was designed to: 1) provide participants with situated experience of this layered interaction and 2) elicit reflection on where meta-level intervention supported or constrained their learning, what challenges arose around the distribution of control, and how they would redesign the interaction.

\subsection{Participants and Study Context}
The study participants were 18 graduate students (17 Master’s students and one PhD student; aged 23-46 years, M = 28.67, SD = 6.93) from a large European technical university who were recruited as knowledgeable co-designers in late Spring 2026. The participants had prior academic backgrounds related to AI, HCI, and learning analytics, as well as ethical considerations related to AI-mediated learning. This background enabled them to reflect both on their own experiences as users of conversational AI systems in their own studies, and broader pedagogical and ethical implications of AI-mediated learning. The study context was the students' use of ChatGPT for course-related learning tasks, where \textit{ChatWise} was introduced to support learners in retaining decision-making authority over their cognitive engagement and interactions with ChatGPT. 

Participation was voluntary, and all workshop attendees received written information about the study aim and procedure and completed a written informed consent form prior to data collection. 

\subsection{Procedure}
\begin{figure}
   \includegraphics[width=1\textwidth]{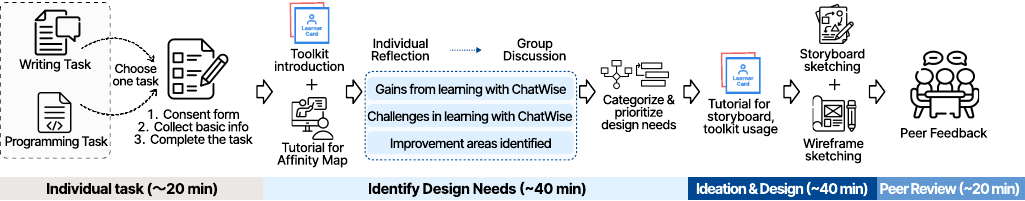} 
    \caption{Workshop Activities (Adapted from Fan et al. \cite{fan2026learneragency})}
    \label{fig: Workshop activities} 
\end{figure}

During the co-design workshop, participants were guided through four activities (Fig. ~\ref{fig: Workshop activities}): (1) Individual Task, (2) Identify Design Needs, (3) Ideation \& Design, and (4) Peer Review. 

To support reflection and co-design, we adopted the \textit{LeAF} toolkit \cite{fan2026learneragency} as a facilitation tool to encourage participants to articulate decision points that they considered important in the layered AI interaction and explore what meta-level support is needed and how such support should be configured at those points. The toolkit consists of six card categories: \textit{Learner} describes learner profiles based on level of domain knowledge, SRL skills, or task confidence (e.g., proficient in domain knowledge, novice in SRL); \textit{Behavior} identifies interaction moments (e.g., formulating prompt, revisiting chat history); \textit{Feedback} provides pedagogical knowledge about feedback levels (e.g., task vs. process), learner needs (e.g., affective vs. cognitive), and feedback stages (feed up vs. feed forward). \textit{Support Type} represents how feedback may be provided (e.g., reflective questions, checklists); \textit{Intervention Policy \& Values} suggests intervention conditions (e.g., immediate or delayed support, learner customization) and design values (e.g., privacy). \textit{Evaluation} provides criteria for evaluating interface designs. Each category also included blank cards to encourage participants to propose alternative perspectives beyond the predefined options.

The toolkit enables participants to construct and evaluate meta-level AI intervention scenarios based on the following structure: When \textit{[Learner]} is doing \textit{[Behavior]}, the system provides \textit{[Feedback]} via \textit{[Support Type]}, under \textit{[Intervention Policy \& Values]}. 

\subsubsection{Individual Task.} \label{individual_tasks}
The workshop began with an individual task to allow participants to experience \textit{ChatWise} in use. After providing informed consent, participants installed the \textit{ChatWise} extension on their laptops. They received a brief introduction to \textit{ChatWise} as a feedback support tool for interacting with commercial conversational AI such as ChatGPT. Participants then performed a learning task using ChatGPT under two conditions, first without \textit{ChatWise} support and then with the \textit{ChatWise} extension enabled, allowing them to directly compare the two learning experiences. They were required to select either a programming or a writing task, see our OSF repository \footnote{ \url{https://osf.io/5b4mg/overview?view_only=5b4fddcefcb64f36a9d2009e8f731324}}. In the writing task, participants selected an empirical research paper and wrote a short paper review, including specific strengths and weaknesses supported by evidence from the paper. In the programming task, participants completed two LeetCode Easy-level programming problems.

\subsubsection{Identify Design Needs.} After the individual task, the facilitator introduced the toolkit, explaining its categories and illustrating them with examples. After that, the participants were divided into six groups to discuss their experience with \textit{ChatWise} in \textit{Miro} \footnote{\url{https://miro.com/}} (i.e., an online collaborative whiteboard tool that helps teams work together visually). Using the card toolkit as inspiration,they responded individually to three questions: (1) What gains did you experience when using \textit{ChatWise} to perform the task? (2) What challenges did you encounter when using \textit{ChatWise} during the task? (3) What could be improved in \textit{ChatWise} to better support your AI-mediated learning? Following that, the participants engaged in group discussions, where they collaboratively used affinity diagramming \cite{lucero2015affinity} to categorize their comments into design themes. They then prioritized design needs based on: (1) 'must' be implemented, (2) 'should' be implemented, and (3) 'could' be implemented.

\subsubsection{Ideation \& Design.} The next phase includes storyboarding and wireframe sketching based on their identified design needs. Participants used the toolkit to elaborate their storyboard individually that shows the improved design. Participants then evaluated each storyboard through individual dot voting within groups. The voting used three criteria: \textit{usefulness, learning impact}, and \textit{feasibility}. Usefulness is explained in terms of whether the storyboard met learners' needs and would improve the interaction experience with conversational AI. Learning impact – whether the proposed requirement could meaningfully influence learners' thinking and the learning process. Feasibility refers to whether the idea could be integrated into conversational AI without major engineering barriers. After selecting storyboard directions, the groups sketched wireframes (simple visual blueprints of the revised \textit{ChatWise} extension) of their revised \textit{ChatWise} concepts.

\subsubsection{Peer Review.} The workshop concluded with oral group presentations and peer critique. Each group shared its design proposal with the larger group. Other participants wrote comments on the proposals. Their reflective comments were structured around three aspects: perceived strengths, opportunities for improvement, and unresolved questions or risks for further exploration. Participants were encouraged to use the \textit{Evaluation Card} to discuss and evaluate the proposed meta-level AI redesigns.

\subsection{Data Analysis} \label{3.3}
Across the 18 participants who provided their student-LLM chatlogs, we obtained 227 logged interaction entries. Each entry contains a learner prompt submitted to ChatGPT and the corresponding meta-level feedback generated by \textit{ChatWise}, which we refer to as a learner prompt-\textit{ChatWise} feedback pair, together with the participant ID, \textit{ChatWise}'s SRL classification label, and a timestamp. For example, one learner prompt-\textit{ChatWise} feedback pair consists of the learner prompt 'can you please first give some possible options?', and the corresponding meta-level feedback asking the learner to clarify the task, describe what they had already considered, and specify what they hoped to achieve. 
Each participant contributed one or more logged interactions. We excluded 42 entries that were not related to the learning tasks, leaving a final total of 185 entries. Of these, 59 were associated with the programming task and 126 with the writing task.

In addition, we collected 174 individual written reflection comments, 17 storyboards, and 5 group design proposals. One group proposal was excluded because a group member did not consent to the use of their data for this research. The storyboards and group design proposals are available in our OSF repository \footnote{ \url{https://osf.io/3gnhq/overview?view_only=20050ce8e45a41c8a043b660363de953}}. 

We adopted a mixed qualitative analysis approach. First, we used deductive coding in chat logs to examine how learners responded to the meta-level AI support (ChatWise) when engaging with the primary AI, namely ChatGPT. We then conducted a reflexive thematic analysis of the coded interaction patterns, participants’ reflections, and design artefacts to gain a deeper understanding of learners’ experiences with layered AI interactions.

\subsubsection{Deductive Coding of Chat Logs}
Learner prompts and corresponding \textit{ChatWise} feedback were first coded separately according to their interactional purpose. We then compared each \textit{ChatWise} intervention with the learner's subsequent prompt to examine whether the subsequent prompt shows observable traces consistent with the meta-level suggestion. We combined the comparison results with the interactional purpose code assigned to the preceding learner prompt to determine under which interactional purposes observable consistency with, or bypass of, the meta-level feedback occurred. In the writing logs, for instance, after \textit{ChatWise} asked one learner to clarify more specifically what they had already considered, the learner's subsequent prompt provided several aspects of the paper that they viewed as strengths or weaknesses. We treated this as observable evidence consistent with the prior meta-level suggestion.

When the learner's subsequent prompt did not show observable traces consistent with the \textit{ChatWise} suggestion, we compared the purpose of the \textit{ChatWise} intervention and the content of the preceding learner prompt to examine whether there were observable factors that could account for the bypass. During this process, we did not attribute learner prompt changes to the offered meta-level AI support, because such changes could also result from the primary AI's response, learners' evolving reasoning, or other aspects of the conversational context. 

\subsubsection{Reflexive Thematic Analysis across Chat-Log Patterns, Reflections and Design Artifacts}
A reflexive thematic analysis \cite{braun2006thematic,braun2019reflecting} was performed on the coded chat-log patterns, the self-reported reflections of the participants, the peer critiques and their design artifacts. This approach was chosen because it is suitable for examining both shared and divergent experiences of layered AI interactions and enables us to explore the underlying assumptions about learner control, authority and decision making.

We first familiarized ourselves with the materials and performed initial coding to capture a broad range of participants' gains, concerns, and their desired system features. During the initial analysis, we identified recurring challenges in learners’ interactions with the primary AI and the meta-level AI, alongside interaction mechanisms that were frequently mentioned in relation to these challenges. Although these challenges relate to the design considerations discussed in Section \ref{2.3}, our analysis revealed how they manifest as competing needs and expectations of meta-level AI support. Therefore, we interpreted them as potential \textit{‘design tensions’} and performed a second round of coding to refine their scope and boundaries. Our analysis resulted in the four design tensions reported in the Findings (see Section \ref{findings}).

\section{Findings} \label{findings}
Our deductive analysis of the learner-LLM logs suggests that learners' subsequent prompts did not always consistently align with the meta-level AI suggestions they received. We identified four observable response patterns: (1) observable responses consistent with its suggestions, (2) continued task progression while bypassing prompted reflection, (3) explicit contestation of the meta-level intervention, and (4) an indeterminate category in which prompt quality observably improved, but it remained unclear whether the improvement stemmed from meta-level AI feedback, primary AI input, or learners’ own reasoning. The first two response patterns occurred in certain interaction contexts. The observable alignment occurred when learners clarified task requirements and context, expressed prior understanding or judgments, outlined solution approaches, or sought further guidance on suggested reasoning activities. Bypass was observed when learners wanted to request complete code or drafts or sought surface-level revisions such as wording or format.


To understand the possible reasons behind learners' different responses to meta-level AI feedback, we conducted a reflexive thematic analysis across the coded chat-log patterns, participants' reflections and peer critiques, and design artifacts. Four themes were developed. We refer to them as \textit{design tensions} in learners' interactions with primary- and meta-level AI: (1) assistance versus reflection, (2) immediate intervention versus space for self-regulation, (3) visibility versus attention competition, and (4) system adaptation versus user control. These tensions highlight learners' challenges related to different consequential decisions. Table \ref{tab:rta_themes} summarizes the subthemes within each tension and indicates the data sources contributing to each subtheme. In the following sections, we use these tensions to organize our findings on how learners experienced and negotiated agency when interacting with a task-oriented primary AI alongside
meta-level AI guidance (\textbf{RQ1}). Within each tension, we also present participants' proposed design responses to these challenges (Group 2-6, see Table \ref{tab:design_proposals}), which inform our discussion of \textbf{RQ2}.  



\begin{table}[t]
\centering
\caption{Themes and subthemes identified through reflexive thematic analysis across chat-log patterns \textbf{[L]}, participants’ reflections on \textit{ChatWise} \textbf{[R]}, and design artifacts \textbf{[A]}.}
\label{tab:rta_themes}
\renewcommand{\arraystretch}{1.0}
\begin{tabular}{p{0.24\textwidth} p{0.72\textwidth}}
\toprule
\textbf{Themes (Design Tensions)} & \textbf{Subthemes} \\
\midrule

\textbf{Assistance vs Reflection} &
Meta-level interventions prompt task-relevant reasoning and reflection \textbf{[R]}. \\[2pt]
&
Reflective intervention may not always fit the learner's immediate interactional purpose \textbf{[L, R]}. \\[2pt]
&
Meta-level AI could misjudge the appropriateness of delegation when earlier conversational turns were not taken into account \textbf{[L, R]}. \\

\midrule

\textbf{Immediate Intervention vs. Space for Self-Regulation} &
Immediate primary-AI assistance can reduce engagement with subsequent reflection \textbf{[R, A]}. \\[2pt]
&
Learners valued reflection that remained available across iterative interaction \textbf{[R, A]}. \\

\midrule

\textbf{Visibility vs Attention Competition} &
Learners reported losing access to meta-level feedback before they were ready to engage or re-engage with it \textbf{[R]}. \\[2pt]
&
Learners experienced prominent or information-dense meta-level feedback as competing for attention with the primary AI interaction \textbf{[R]}. \\[2pt]
&
Learners sought control over when meta-level feedback remained peripheral or became focal \textbf{[R, A]}. \\

\midrule

\textbf{System Adaptation vs User Control} &
Learners wanted to retain control over how meta-level support was personalized and updated \textbf{[R, A]}.\\[2pt]
&
Learners sought meta-level support adapted to their expertise and task context \textbf{[R, A]}.\\

\bottomrule
\end{tabular}
\end{table}

\begin{table*}[t]
\centering
\caption{Overview of the five group design proposals and their key interaction features.}
\label{tab:design_proposals}
\renewcommand{\arraystretch}{1.15}
\setlength{\tabcolsep}{5pt}

\begin{tabularx}{\textwidth}{
    >{\raggedright\arraybackslash}p{0.1\textwidth}
    >{\raggedright\arraybackslash}X
    >{\raggedright\arraybackslash}p{0.30\textwidth}
}
\toprule
\textbf{Group} & 
\textbf{Design proposal} 
\\
\midrule

Group 2 &
Proposed a guided prompt-refinement workflow in which a broad or vague prompt triggers a small feedback card. Learners can dismiss the intervention or choose to improve the prompt, after which \textit{ChatWise} provides structured, step-by-step guidance. The design also allows learners to configure the detail, tone, and format of feedback and to respond to \textit{ChatWise} when the guidance does not fit their needs. 
\\

\midrule

Group 3 &
Proposed relocating \textit{ChatWise} feedback to a dedicated side region so that meta-level guidance remains visually distinct from the primary ChatGPT conversation and does not repeatedly obstruct the ongoing interaction. 
\\

\midrule

Group 4 &
Proposed providing feedback before a learner submits a prompt to ChatGPT. \textit{ChatWise} offers selectable suggestions for revising the prompt, allowing learners to inspect and incorporate guidance before engaging the primary AI. 
\\

\midrule

Group 5 &
Proposed a persistent and learner-configurable \textit{ChatWise} space combining concise feedback with retained feedback history. Learners can revisit earlier feedback, indicate their experience level, and use preference controls such as like/dislike to shape subsequent support. 
\\

\midrule

Group 6 &
Proposed a Grammarly-like interaction in which \textit{ChatWise} remains lightweight by default through a small toggle or inline indicator, while more detailed guidance can be expanded when needed. Feedback is presented as concise bullet points or checklists and is designed to be content-aware, aligned with learner intent, and actionable. The workflow can also guide learners through structured selections before generating an improved prompt. 
\\

\bottomrule
\end{tabularx}

\vspace{3pt}
\parbox{\textwidth}{%
    \footnotesize
    \textit{Note.} Group 1 was excluded because one group member
    did not consent to the use of their data in this study.
}
\end{table*}

\subsection{Tension 1: Assistance vs Reflection} \label{4.1}
Our analysis revealed a tension between the support offered by the primary and meta-level AI. The primary AI was valued for reducing task effort through direct assistance, while the meta-level intervention was found helpful in encouraging learners to remain reflective and cognitively engaged. Participants reported both benefits and frustration with the meta-level AI support depending on their intentions during the learning task. 

Participants in both task groups described \textit{ChatWise} as prompting more deliberate engagement. W2 \footnote{W and P denote participants in the writing and programming tasks, respectively.} reported that it helped them ``snap out of autopilot mode'' and reflect on ``what I am actually asking and what I want to learn from the task,'' while P3 expressed being reminded to ``try with my knowledge at first.'' The cognitive work that the participants sought to retain varied across the two tasks. In the setting of the programming task, several participants associated reflective engagement with problem-solving and independent attempts (n = 3). Whereas in the context of the writing task, the participants focused more on articulating the writing intention and refining the scope and depth of prompts (n = 5).  

However, the participants did not necessarily consider additional reflection useful for every interaction with ChatGPT. W9, for example, stated, ``I don’t want the prompt to be perfect every time.'' The student-LLM logs further show that learners wrote prompts not only to address task- or reasoning-related problems, but also to manage ongoing dialogue with the primary AI, such as by acknowledging understanding (n = 3) or supplementing earlier requests to adjust response length or format (n = 2). This suggests that reflective interventions may not always align with learners' immediate interactional purposes.

Participants also questioned \textit{ChatWise}'s judgments regarding ``appropriate delegation''. Several participants expressed frustration and questioned interventions that treated what they considered to be appropriate delegation as a problem. After completing the programming task, P7 complained that they had already finished their solution and only wanted ChatGPT to assist with a smaller formatting error, but \textit{ChatWise} still issued a red warning against earlier flagged 'direct assistance'. Another participant (P8) reported that they followed \textit{ChatWise}'s advice to break a problem into smaller steps, but subsequent stepwise requests were still treated as overly direct. In response to these concerns, P8 suggested that \textit{ChatWise} needs to recall its earlier recommendations when generating subsequent feedback and allow users to respond directly to its guidance. Similarly, P1 asked for greater awareness of the conversational context without requiring learners to repeat earlier instructions.

From an agency perspective, participants' accounts highlight challenges for meta-level AI in supporting \textbf{cognitive agency}. This concerns learners' control over which cognitively consequential activities they undertake themselves and which they delegate to the primary AI. 
Our findings indicate that supporting cognitive agency is not merely a matter of reducing direct AI assistance or imposing stronger reflective constraints. 
Rather, effective reflective intervention requires considering learners' intentions, including their own judgments about which cognitive activities to undertake themselves or delegate to AI, and whether additional reflection fits their immediate interactional purpose. This, in turn, requires considering the conversational and learning context.

\subsection{Tension 2: Immediate Intervention vs Space for Self-Regulation} \label{4.2}

Participants' reflections reveal different expectations about the timing of the reflective meta-level intervention. Some comments emphasize the value of a reflective pause before receiving assistance from the primary AI (n = 5), while other comments highlight the importance of leaving room for learners to develop their task understanding and reflect on their approaches across successive interactions (n = 2).

Three participants explicitly expressed reduced motivation to engage with \textit{ChatWise} after seeing ChatGPT had already produced a ``ready-made solution.'' W3 explained that ``ChatGPT already gave me the answer so I could easily ignore \textit{ChatWise}.'' This concern was also reflected in the participants’ redesigns of ChatWise: 10 of the 17 storyboards placed meta-level intervention during prompt formulation before executing the actual prompt (see the storyboards in our OSF repository \footnote{ \url{https://osf.io/3gnhq/overview?view_only=20050ce8e45a41c8a043b660363de953}} ). 

However, the study participants did not necessarily envision \textit{ChatWise}'s reflective support as a one-shot checkpoint in which \textit{ChatWise} should immediately resolve how the learner ought to proceed. Some participants instead described 'task understanding' and 'prompt formulation' as iterative processes that unfolded across multiple turns. One student (W1) in the writing group proposed to ``optimize task understanding through multiple prompts in a chat'' and to progressively ``dial in'' prompt quality. This suggests that reflective engagement could develop through continuous conversation rather than a single corrective intervention. W2 also noted that \textit{ChatWise }'s feedback was not accessible after some seconds, and that they ``could not come back to it when thinking about [their] next prompt,'' indicating that the AI-mediated metacognitive guidance could remain relevant beyond the moment it was first delivered. 

Taken together, these findings highlight a temporal tension between interrupting the immediacy of primary-AI assistance to create space for reflection and maintaining that reflection as the interaction progresses. From an agency perspective, this tension relates to \textbf{Task Agency}, as the temporal ordering of meta-level intervention and primary-AI responses can shape learners' decisions about how to proceed, for example, whether to revise or continue with their current task plan. It also affects \textbf{Cognitive Agency} allocation, influencing whether underlying planning and reflection remain learner-led or are replaced by complete AI responses or overly prescriptive meta-level guidance.


\subsection{Tension 3: Visibility vs Attention Competition} \label{4.3}
Participants' reflections and design proposals highlight the need for meta-level feedback to remain noticeable enough to inform learners' ongoing reflection and interaction. Yet making meta-level guidance more prominent could also increase its attentional demand, creating competition with the primary AI and disrupting the learner's ongoing task flow.

P3 noted that the \textit{ChatWise} mediated feedback ``disappeared too quickly'' while they were still focused on solving the task, and P1 proposed to ``keep feedback on screen but allow conversations [with ChatGPT] to continue as well.''. At the same time, several participants described various forms of attentional interference. \textit{ChatWise }could ``cover up the prompt window'' (P1), obscure the ChatGPT response (W8, P4), ``suddenly appear'' in a floating window that ``interrupted my train of thought and focus'' (P5). Others reported wanting to ignore feedback when they perceived it as overly lengthy (W1, W3, P2, P9), suggesting that attentional competition arose not only through spatial obstruction but also through information load. 

\begin{figure}
   \includegraphics[width=0.51\textwidth]{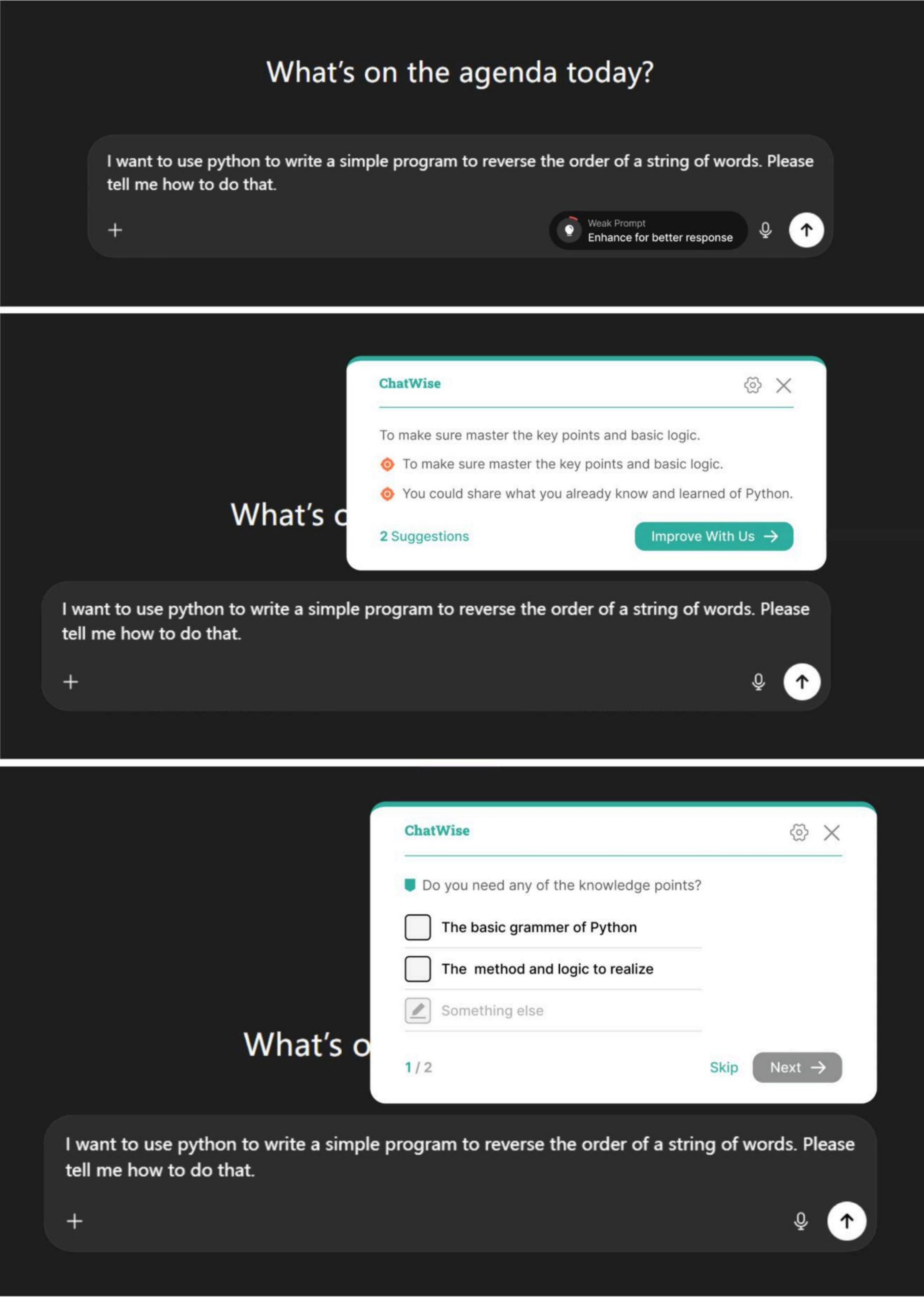} 
    \caption{Group 6's design proposal with an inline prompt indicator and expandable feedback panel.}
    \label{fig:Group 6's design proposal}
\end{figure}

However, participants did not respond by simply minimizing \textit{ChatWise.} Instead, in their design proposals, they sought ways for \textit{ChatWise} to remain available without continuously occupying focal attention or presenting all of its guidance at once. Several groups proposed progressive disclosure, presenting meta-level guidance in a lightweight form by default—through small indicators, minimized cards, or concise summaries—and allowing learners to reveal more detailed feedback on demand (see Table \ref{tab:design_proposals}). Group 6, for example, paired a minimal inline indicator with an expandable feedback panel (Fig.~\ref{fig:Group 6's design proposal}), while Group 2 presented an initial lightweight feedback card that allowed learners to decide whether to receive further guidance on improving their prompt. Group 5's proposals complemented this with dedicated interface regions or retained feedback histories that kept guidance accessible without requiring continuous focal attention. These design ideas sought to coordinate the attentional roles of the two AI layers. These design ideas that enable learners to expand on, dismiss or revisit guidance from \textit{ChatWise} offer insights into how meta-level AI can support \textbf{Interaction Agency} by giving learners greater control over whether the meta-level guidance remains peripheral or became the focus of engagement.



\subsection{Tension 4: System Adaptation vs User Control}
\label{4.4}

Most participants wanted meta-level guidance to adapt automatically according to their knowledge expertise and task context, while still enabling them to control how this adaptation is configured and updated over time. We identified this tension when participants first criticized the limitations of generic, context-insensitive guidance and then suggested design ideas that gave users more explicit ways to configure, correct, and update how guidance was personalized.

Across the two task groups, participants highlighted different aspects of how \textit{ChatWise} 'failed' to accommodate their expertise and support needs. Participants in the programming group emphasized expertise and prior knowledge adaptation more often. P1 reported, for example, that the feedback provided by \textit{ChatWise} was ``very base-level/generic'' and argued that it should ``take into account level of experience'', explaining, ``I dont need to/want to explain what a set is in my prompt and why it is a good idea.'' P5 also requested a mastery-based knowledge verification, for example, by asking, ``Do you understand the Programming Syntax?'' Writing groups, by contrast, more often focused on how feedback was delivered, suggesting wording that was ``more encouraging and motivating'' (W3) or ``more natural'' and closer to feedback ``given by an actual person'' (W8).

In response, the participants proposed several ways for learners to steer how \textit{ChatWise} can adapt its support (see Table \ref{tab:design_proposals}). Group 5 introduced an experience-level setting through which learners could indicate their prior knowledge, alongside 'like/dislike' controls intended to help the system learn what kinds of feedback they found useful over time. Group 2 similarly proposed configurable settings for feedback detail, tone, and format, with preferences that could be updated over time. W6 also suggested that \textit{ChatWise} could ``store contextual information from previous tasks with the user's consent, similar to ChatGPT.'' These proposals offer insights into how meta-level AI can support \textbf{Interaction Agency}, particularly in terms of participants’ desire for greater authority over how \textit{ChatWise} adapted its support to their expertise and learning context, as well as over what information could be retained and reused to inform future adaptation.

\section{Discussion}


This study examined how learners experience and negotiate agency when reflective meta-level AI is introduced alongside a task-oriented generative AI system. Our findings identified four interconnected design tensions: (1) assistance versus reflection, (2) immediate intervention versus space for self-regulation, (3) visibility versus attention competition, and (4) system adaptation versus user control. They also revealed how learners attempted to address these tensions through different interaction-design proposals.

Rather than suggesting that agency can be preserved simply by increasing reflection or providing additional interface controls, our findings indicate that agency depends on which decisions remain consequential for learners and how authority over these decisions is distributed between the learner, primary AI, and meta-level AI. We discuss the implications for conceptualizing learner agency, designing reflective scaffolding, and developing agency-preserving AI interactions.

\subsection{Learner Agency as Negotiated Decision Authority across Layered AI Interaction} \label{5.1}

Our findings extend understanding of learner agency in AI-mediated learning by distinguishing three interrelated dimensions of decision authority: task agency, cognitive agency, and interaction agency.

Existing HCI research increasingly conceptualizes agency as relational and dynamically configured through human-AI interaction rather than as an individual attribute or a fixed distribution of control \cite{bennett2023agency, zhang2025collaboration}. Similarly, the Agency Allocation Framework proposed by Borchers, Viberg and Kizilcec \cite{borchers2026agency} emphasizes the importance of examining specific decision points, the actors involved, and their respective authority. Our findings build on these perspectives by illustrating how decision authority becomes distributed not only between a learner and a task-oriented AI, but also across a secondary AI system intended to regulate the learning interaction.

\textit{Task agency} concerns learners' authority over the direction, sequencing, and progression of their learning activities. Our findings suggest that this dimension becomes particularly consequential when meta-level interventions influence when learners proceed with a task, revise their prompts, or receive assistance from the primary AI. Participants' preference for guidance before prompt submission illustrates how the temporal organization of AI support may influence these decisions. However, participants also emphasized the importance of iterative interaction, suggesting that preserving task agency requires maintaining opportunities to revise and reconsider task strategies rather than enforcing a predetermined sequence of reflective steps. This distinction builds on Chu-Carroll and Brown's differentiation between task initiative and dialogue initiative \cite{chucarroll1997tracking}. While conversational systems may provide opportunities for users to direct the immediate dialogue, this does not necessarily imply that they retain authority over the broader task trajectory. Our findings extend this distinction to 'layered AI interactions', where a meta-level system may influence task progression without directly performing the learning task itself.

\textit{Cognitive agency} concerns learners' authority over which reasoning, planning, evaluation, and problem-solving activities they undertake themselves and which they delegate to AI. Prior research has highlighted the risk that generative AI can facilitate cognitive offloading by making complete solutions readily available \cite{jelson2026empirical, isa2026codegeneration}. Taylor and Krishna's \cite{taylor2026contextmaxxing} conceptualization of cognitive agency similarly foregrounds users' capacity to exercise judgment, monitor their thinking, and develop understanding through AI collaboration.

Our findings add an important qualification: preserving cognitive agency does not necessarily require minimizing cognitive delegation. Participants considered some forms of AI assistance appropriate, particularly when they had already completed the cognitively demanding parts of a task or sought help with relatively superficial revisions. Conversely, meta-level interventions that failed to recognize these circumstances were experienced as unnecessary or inappropriate. This suggests that cognitive agency depends not simply on the amount of reasoning delegated, but on learners' opportunities to make informed and contextually meaningful decisions about that delegation.

\textit{Interaction agency} concerns learners' authority over how AI systems participate in their learning, including the timing, visibility, intensity, and personalization of support. Participants sought opportunities to dismiss, expand, revisit, configure, and contest meta-level feedback. These preferences align with research emphasizing that meaningful agency requires more than nominal access to choices; learners must be able to influence subsequent system behavior \cite{pauzi2026evaluating}. The distinction between these three dimensions contributes to existing conceptualizations of learner agency by making different objects of decision authority analytically visible. Importantly, the dimensions are interdependent. For example, an intervention designed to strengthen cognitive agency by encouraging independent reasoning may simultaneously restrict task agency by delaying progression and interaction agency by limiting the learner's ability to dismiss the intervention.

We therefore propose that agency in layered AI environments should be understood as a negotiated configuration of decision authority across multiple dimensions, rather than as a single property that can be uniformly increased through additional scaffolding or user controls. 

\begin{table*}[t]
\caption{Interaction mechanisms for Agency-Preserving Meta-level AI.}
\label{tab:interaction_mechanisms}
\centering
\scriptsize
\setlength{\tabcolsep}{5pt}
\renewcommand{\arraystretch}{1.15}

\begin{tabularx}{\textwidth}{
    >{\raggedright\arraybackslash}p{0.12\textwidth}
    >{\raggedright\arraybackslash}p{0.32\textwidth}
    >{\raggedright\arraybackslash}p{0.38\textwidth}
    Y
}
\toprule
\textbf{Mechanism} &
\textbf{Explanation} &
\textbf{Agency-related Design Challenges} &
\textbf{Connection to Findings} \\
\midrule

\textbf{Stepwise Cognitive Scaffolding} &
Breaking broad reflective guidance into a sequence of structured and concrete cognitive prompts that help learners work through reasoning, such as articulating their intent or decomposing a problem. &
\textbf{Cognitive Agency:}
How can meta-level guidance structure complex reflection into actionable steps that sustain learners' active participation in reasoning over time? &
See Section~\ref{4.1} \\

\midrule

\textbf{Context-aware Assessment} &
Assessing whether intervention is warranted by considering what the learner has already done, what they are currently delegating to the primary AI, and which reasoning activities remain important for the learner to perform. &
\textbf{Cognitive Agency:}
How should the meta-AI interpret the learner's current situation before deciding whether an intervention is warranted and what it should target? &
See Section~\ref{4.1} \\

\midrule

\textbf{Learner Correction and Contestability} &
Making the meta-level AI's assessment and resulting intervention open to learner clarification, challenge, or correction when they do not reflect the learner's prior work, task state, or intended delegation. &
\textbf{Cognitive Agency:}
How can learners challenge or correct the meta-AI's judgment about which reasoning activities should remain learner-led?
\vspace{1pt}

\textbf{Interaction Agency:}
How can learners contest, redirect, or override an intervention when it does not fit their current situation? &
See Section~\ref{4.1} and Section~\ref{4.4}  \\

\midrule

\textbf{Cross-turn Memory} &
The meta-AI should retain awareness of its previous guidance and the learner's prior attempts across turns. &
\textbf{Cognitive Agency:}
How can meta-level AI carry forward learners' prior reasoning and earlier guidance so that cognitive work already undertaken remains recognized in subsequent judgments, while avoiding repetitive or contradictory interventions? &
See Section~\ref{4.1} \\

\midrule

\textbf{Guided Prompt Formulation} &
Integrating meta-level guidance into prompt formulation, where learners are already clarifying, framing, and refining their requests, so that this reasoning can be supported before primary-AI assistance takes over the task interaction. &
\textbf{Task Agency:}
How can meta-level guidance preserve learners' opportunity to reconsider and revise how they want to proceed before primary-AI assistance advances the task?
\vspace{1pt}

\textbf{Cognitive Agency:}
How can meta-level guidance support the reasoning involved in prompt formulation without taking over the cognitive work that learners could perform themselves? &
See Section~\ref{4.2} \\

\midrule

\textbf{Iterative Feedback Loops} &
Reflection is supported across multiple turns rather than treated as a one-shot checkpoint. Learners can progressively refine their understanding and approach during the conversation. &
\textbf{Task Agency:}
How can iterative meta-level feedback preserve learners' authority to determine the direction of subsequent actions as the task progresses?
\vspace{1pt}

\textbf{Cognitive Agency:}
How can iterative feedback keep learners cognitively involved as their understanding develops across turns? &
See Section~\ref{4.2} \\

\midrule

\textbf{Feedback and Revision Traceability} &
Maintaining an accessible trace of earlier meta-level feedback and subsequent revisions so that learners can revisit prior guidance as they plan and refine later actions. &
\textbf{Task Agency:}
How can prior feedback and revisions remain available to inform learners' decisions about how to proceed as the task develops?
\vspace{1pt}

\textbf{Cognitive Agency:}
How can traces of prior feedback and revision remain available for learners to reconsider and build on their earlier reasoning? &
See Section~\ref{4.2} \\

\midrule

\textbf{Peripheral-to-Focal Transition} &
Meta-level guidance remains lightweight and peripheral by default, while allowing learners to choose when to bring it into focal attention for more detailed engagement. &
\textbf{Interaction Agency:}
How can meta-level guidance remain available without demanding continuous focal attention, while allowing learners to decide when it should become the focus of engagement? &
See Section~\ref{4.3} \\

\midrule

\textbf{User-Controlled Re-engagement} &
After meta-level guidance leaves focal attention, learners can deliberately bring it back when it becomes relevant again. &
\textbf{Interaction Agency:}
How can learners retain authority over whether and when to re-engage with earlier meta-level guidance? &
See Section~\ref{4.3} \\

\midrule

\textbf{Spatial Separation of AI Roles} &
Structuring the interface to make the primary AI and meta-level AI spatially and functionally distinguishable, so that learners can differentiate their roles without the two layers competing for the same attentional space. &
\textbf{Interaction Agency:}
How can the spatial organization of primary- and meta-AI roles reduce interface-imposed competition while allowing learners to determine which AI layer to engage with? &
See Section~\ref{4.3} \\

\midrule

\textbf{Expertise-Sensitive Adaptation} &
Adapting meta-level guidance to learners' existing knowledge and experience, so that the scope, level, and form of support can better align with their current expertise. &
\textbf{Interaction Agency:}
How can meta-level AI adapt support to learners' existing expertise without imposing guidance that is misaligned with what they already know? &
See Section~\ref{4.4} \\

\midrule

\textbf{User-Steerable Personalization} &
Giving learners explicit ways to configure and update how meta-level guidance is personalized. &
\textbf{Interaction Agency:}
How can learners retain authority over how meta-level guidance is personalized, including which preferences and learner information shape future adaptation? &
See Section~\ref{4.4} \\

\bottomrule
\end{tabularx}
\end{table*}

\subsection{Designing Meta-Level AI That Learners Can Regulate: Criteria and Interaction Mechanisms}

This section concerns the interaction-design requirements for reflective meta-level AI. Prior work has demonstrated the potential of metacognitive scaffolding to support learners' reflection, monitoring, and evaluation of AI-generated information \cite{gmeiner2025metacognitive, park2026llmbox, wang2025srlmentor}. However, our findings suggest that the educational value of such interventions depends not only on their reflective content but also on how they are integrated into learners' ongoing interaction with the primary AI. If agency is a configuration of decision authority rather than a property to be increased (Section \ref{5.1}), design cannot simply add reflective prompts or interface controls. It must specify which decisions learners retain authority over and how.

Across the four design tensions, participants proposed different approaches to improving the interaction between task-oriented and meta-level AI. Table \ref{tab:interaction_mechanisms} summarizes the interaction mechanisms identified across the four design tensions, illustrating how learners sought to negotiate decision authority through concrete features of meta-level AI support. Although these proposals varied in their specific features and implementation strategies, they reflected several recurring concerns about learners' ability to influence when, how, and under what conditions metacognitive guidance should be provided.

By examining the proposals in relation to the reported challenges, participant reflections, and peer critiques, we synthesized \textit{five overarching design criteria for agency-preserving meta-level AI} \textit{(DC1-DC5)}. The synthesis does not suggest all proposed features were equally desirable or that increasing learner control is sufficient in itself. Instead, we considered how the design proposals addressed the identified tensions and whether their underlying mechanisms could support reflective engagement without unnecessarily restricting learners' decision-making authority. Each criterion brings together related interaction mechanisms (see Table \ref{tab:interaction_mechanisms}), which differ in the dimension of agency they primarily concern.

\textit{DC1. Context-sensitive and contestable guidance.} 
Prior research suggests that learners' decisions about cognitive delegation vary according to their goals, expertise, and task context \cite{song2026goodfences}. Our findings illustrate the implications of this variability for reflective support in layered AI interaction. Participants questioned feedback that interpreted legitimate assistance requests as inappropriate delegation or failed to recognize earlier reflective efforts. These experiences indicate that reflective interventions should consider conversational history and the learner's immediate interactional purpose with the primary AI, as well as the meta-level AI's own earlier guidance. Importantly, contextual adaptation should be accompanied by opportunities to correct or contest the system's interpretation. Otherwise, the meta-level AI risks becoming the sole authority for determining what constitutes appropriate learning behavior. This criterion recognizes that judgments about appropriate cognitive delegation should not be determined exclusively by the meta-level AI.


\textit{DC2. Learner-controlled timing of reflection.} 
Existing research suggests that combining immediate and delayed feedback may support both reflection-\textit{in}-action and reflection-\textit{on}-action \cite{zhang2026its}. Our findings further indicate that, in layed AI interaction, the timing of reflection is consequential because primary AI responses can rapidly change learners' motivation to engage with subsequent guidance. The preference for pre-submission feedback in many participant storyboards suggests that reflection may be more meaningful when learners can consider guidance before receiving complete assistance. However, the findings also suggest that reflection can develop iteratively rather than through a single intervention. The value of pre-submission support should not be equated with mandatory intervention. Instead, systems should create opportunities for reflection at consequential decision points while allowing learners to proceed, defer, or revisit the guidance. Learners should thus retain control over the timing of reflective engagement. 


\textit{DC3. Persistent and revisitable feedback.} 
Existing approaches to SRL scaffolding often distinguish immediate diagnostic support from later opportunities for reflection \cite{wang2025srlmentor}. Our findings suggest that feedback persistence may provide another way to support this temporal flexibility. Participants described reflective suggestions that became relevant only after subsequent task interactions, but were no longer accessible. Maintaining feedback histories and revision traces may allow learners to reconnect earlier guidance with their evolving understanding. This shifts the design emphasis from delivering feedback at an assumed optimal moment toward preserving opportunities for learner-controlled re-engagement. Learners can then determine when previously delivered feedback becomes relevant to their ongoing reasoning and task decisions.


\textit{DC4. Peripheral-to-focal presentation.} 
The attentional demands of reflective support have been recognized in prior research on information overload and the presentation of AI-generated explanations \cite{li2026supporting, cheng2025noticeability}. Our findings connect these concerns to learners' decision authority. Participants did not simply request less feedback; rather, they proposed ways to control \textit{when }feedback should occupy focal attention. Their proposals for expandable cards, dedicated interface regions, and concise indicators suggested a common design direction: guidance should remain accessible without continuously demanding attention. These proposals resonate with the 'calm technology' principle, suggesting that information should move between the center and periphery of attention \cite{weiser1996calm}. In layered AI environments, progressive disclosure and spatial separation can therefore serve not only usability goals but also agency-related purposes by allowing learners to regulate the attentional relationship between primary and meta-level AI. This criterion emphasizes the learners' ability to regulate attentional prominence.


\textit{DC5. Learner-steerable adaptation.} 
Research on AI-mediated learning has emphasized the importance of personalization in responding to differences in learner characteristics and support needs \cite{wu2026afterinterface}. However, adaptation introduces a further question of decision authority: who determines which learner characteristics matter, and what support should follow from them? Participants' proposals for expertise settings, feedback preferences, and corrective responses suggest that personalization should not operate exclusively through automated inference. Instead, it should remain transparent and enable learners to communicate, inspect, and revise relevant assumptions. Such mechanisms may help reconcile the benefits of adaptive support with the need to preserve learners' authority over their learning interactions. This criterion thus emphasizes learner authority over personalization

Taken together, these design criteria extend earlier research on reflective scaffolding to layered AI interaction by shifting attention from the provision of metacognitive guidance to the interaction conditions under which that guidance can remain meaningful, revisable, and subject to learner judgment. They should be understood as design propositions requiring subsequent implementation and rigorous evaluation, rather than the evidence that particular interface features necessarily improve learner agency and ultimately learning in AI-mediated learning settings.

\subsection{The Paradox of Meta-Level AI: Supporting Regulation without Taking Over}

A broader theoretical implication of this study's findings concerns the paradox of introducing additional AI support to preserve learner agency. Research on AI-mediated learning has increasingly explored approaches that introduce deliberate friction, reflective prompts, and pedagogical constraints to counteract excessive cognitive offloading. For example, Klein et al. \cite{klein2026answer} argue that frictionless AI assistance can undermine sustained cognitive engagement and propose stronger structures for preserving educationally meaningful reasoning. Similarly, Sharma et al. \cite{sharma2026hybrid} and Li et al. \cite{li2026supporting} investigated ways of shaping AI interactions to promote more reflective and critical engagement (for more, see Background).

Our findings suggest that such interventions introduce a second-order problem of regulation: \textit{when an AI system is introduced to regulate learners' engagement with another AI system, who regulates the regulator?}

The meta-level AI does not merely provide additional information. Through decisions about when to intervene, what behavior to encourage, which forms of delegation to discourage, and how to personalize feedback, it can influence the conditions under which learners exercise their own judgment. Consequently, an intervention intended to strengthen cognitive agency may unintentionally restrict task or interaction agency. This tension was evident in participants' responses to \textit{ChatWise.} Some appreciated being encouraged to reconsider their prompts and engage more deliberately with learning tasks. Others questioned the system's judgments, particularly when it discouraged assistance that they considered appropriate. Similarly, participants valued adaptive guidance but wanted opportunities to correct or influence how the system interpreted their expertise and preferences. These findings challenge the assumption that more reflective scaffolding necessarily leads to greater learner agency. They also suggest that agency-preserving design cannot be reduced to a simple choice between stronger AI regulation and unrestricted learner autonomy. Stronger constraints may protect opportunities for meaningful reasoning, but they may also become inappropriate when applied without sufficient sensitivity to learners' goals, prior efforts, and task contexts.

Building on the Agency Allocation Framework \cite{borchers2026agency}, we argue that reflective AI support should be examined in terms of the specific decisions it influences and the authority learners retain over those decisions. This includes the ability to accept, reject, defer, revise, or contest interventions, as well as the ability to influence how future support is configured. From this perspective, agency-preserving meta-level AI should not aim to maximize reflection or minimize delegation, but to support learners' capacity to regulate their own engagement with AI assistance. This distinction is particularly relevant to layered AI systems, where the introduction of additional agents or feedback mechanisms can increase the complexity of decision-making. The design challenge is therefore not only to coordinate the functions of multiple AI systems, but also to ensure that their coordination does not make learners' own decisions less consequential.

\subsection{Limitations and Future Research}

Several limitations should be considered when interpreting these findings. \textit{First}, the study involved 18 graduate students with backgrounds related to AI, HCI, and learning analytics. Their familiarity with AI-supported learning and design concepts may have enabled detailed reflection on interaction mechanisms, but their perspectives may not represent those of learners with less technical experience or different educational backgrounds. \textit{Second,} participants engaged in relatively short programming and academic writing tasks within a co-design workshop. Agency-related tensions may unfold differently during prolonged learning activities, where learners develop more stable interaction habits and where the consequences of cognitive delegation become more apparent over time. \textit{Third, }although the chat-log analysis identified observable patterns of alignment, bypass, and contestation, these patterns cannot establish whether meta-level feedback caused changes in learners' behavior. Learners' subsequent prompts may also have been influenced by primary AI responses, prior knowledge, or evolving task understanding.
\textit{Finally,} the proposed design criteria emerged from participants' experiences and co-design artifacts rather than from an evaluation of implemented alternatives. Their effectiveness in preserving learner agency, improving self-regulation, or supporting learning outcomes remains to be established.

Future research should therefore investigate how the proposed criteria can be operationalized in functioning meta-level AI systems and evaluated across different tasks, learner populations, and durations of use. Particularly important questions concern how learners exercise control over intervention timing, how they contest inappropriate guidance, and whether opportunities for learner-steerable adaptation translate into meaningful influence over subsequent system behavior. Longitudinal and experimental studies could further examine the relationships between perceived agency, observable decision-making, cognitive engagement, and learning outcomes. Such work would help establish whether agency-preserving interaction mechanisms can support educationally valuable reflection without introducing unnecessary constraints on learners' autonomy and task progression.

\section{Conclusion}

As generative AI becomes increasingly integrated into learning, preserving learner agency requires more than encouraging reflection or limiting cognitive delegation. This study
examined how learners experience and negotiate decision authority when task-oriented generative AI operates alongside reflective meta-level AI. Through an analysis of learner--AI interactions, participant reflections, and co-design artifacts, we identified four interconnected design tensions concerning assistance and reflection, intervention timing, attention, and adaptation. These tensions reveal a fundamental paradox: AI systems introduced to support learners' self-regulation may themselves constrain agency by determining when, how, and under what conditions learners should reflect.

Our findings contribute to HCI research by demonstrating that learner agency in layered AI interactions is multidimensional and dynamically negotiated across task, cognitive, and interaction agency. Importantly, supporting one dimension of agency may constrain another, highlighting the need to move beyond treating learner control or reflective
engagement as uniformly beneficial. Building on these insights, we synthesized five empirically grounded design criteria, supported by concrete interaction mechanisms,
for developing context-sensitive, temporally flexible, revisitable, attention-aware, and learner-steerable meta-level AI. These criteria provide directions for designing reflective support that remains responsive to learners' goals while preserving opportunities to question, revise, and contest AI interventions.

Ultimately, we argue that the central challenge of agency-preserving AI is not simply deciding how much assistance or reflection AI should provide, but
\textit{who has the authority to make those decisions}. Rather than regulating learners toward predefined forms of desirable engagement, meta-level AI should support
learners in exercising informed judgment over their own cognitive participation, task progression, and AI interactions. The future of agency-preserving AI therefore depends not only on designing systems that help learners reflect, but on ensuring that learners retain meaningful authority over how AI shapes their thinking and learning.

\begin{acks}
This work was supported by KTH Royal Institute of Technology through the project ``Think for Yourself: Adaptive AI Technologies for Learning with LLMs in STEM Education.''
\end{acks}

\section*{Author Contributions}
CRediT: \textbf{Yueling Fan:} Conceptualization, Data curation, Formal analysis, Investigation, Methodology, Resources, Project administration, Visualization, Writing -- original draft, Writing -- review \& editing; \textbf{Jacqueline Wong:} Methodology, Supervision, Writing -- review \& editing; \textbf{Olov Engwall:} Supervision, Writing -- review \& editing; \textbf{Olga Viberg:} Conceptualization, Formal analysis, Funding acquisition, Investigation, Methodology, Project administration, Supervision, Validation, Writing -- original draft, Writing -- review \& editing.

\section*{Disclosure Statement}
The authors report no competing interests. ChatWise, the meta-level AI system examined in this study, was developed by members of the research team for research purposes and is not commercially available.

\section*{Ethics Statement} 
This study was conducted in accordance with established research ethics principles and applicable Swedish regulations. Formal ethical approval was not required under the Swedish Ethical Review Act (2003:460), as the study involved a co-design workshop with university students and did not involve the processing of sensitive personal data, physical or psychological interventions intended to affect participants, or research methods posing an obvious risk of harm. Participation was voluntary, and written informed consent was obtained from all participants included in this study prior to data collection. 

\section*{Declaration of Generative AI Use}
The authors used AI tools including ChatGPT (GPT-6 Astra, OpenAI) and Grammarly for three purposes:

\textit{Summarizing design artifacts.} ChatGPT was used to produce initial summaries of the design features described in participants' group design proposals. The authors thoroughly checked all summaries against the original prototypes and design descriptions and revised them to reflect participants' original design intentions.

\textit{Supporting data analysis without generating codes or themes.} During the initial deductive coding of the chat logs and reflexive thematic analysis, ChatGPT was used to help the authors familiarise themselves with the dataset, locate relevant passages and check that no relevant information had been overlooked and that codes were applied appropriately. ChatGPT was also prompted to propose alternative interpretations that challenged the authors' emerging findings. In such cases, the authors returned to the relevant data to re-examine their interpretations. ChatGPT was not used to generate codes or themes, and all coding and analytical decisions were made by the authors.

\textit{Drafting and language checking.} During initial drafting, the first author used ChatGPT to translate fragmented mixed English-Chinese notes into English in order to record thoughts and ideas quickly. All content was reviewed and edited based on the author's original ideas. The first author then revised the draft 2 to 4 times before preparing the manuscript in Overleaf, and all co-authors reviewed and edited its content. The Grammarly extension was used for grammar and spelling checks in Overleaf. All AI-assisted wording was reviewed sentence by sentence by the authors, and all text reflects the authors' own arguments and interpretations.

\section*{Data Availability Statement}
The task materials, storyboards, and group design proposals that support the findings of this study are openly available on the Open Science Framework (OSF) at
\begin{enumerate}
  \item task materials: \url{https://osf.io/5b4mg/overview?view_only=5b4fddcefcb64f36a9d2009e8f731324}
  \item storyboards, and group design proposals \url{https://osf.io/3gnhq/overview?view_only=20050ce8e45a41c8a043b660363de953}
\end{enumerate}

\bibliographystyle{ACM-Reference-Format}
\bibliography{sample-base}

@inproceedings{lucero2015affinity,
  author    = {Andr{\'e}s Lucero},
  title     = {Using Affinity Diagrams to Evaluate Interactive Prototypes},
  booktitle = {Human-Computer Interaction -- INTERACT 2015},
  series    = {Lecture Notes in Computer Science},
  volume    = {9297},
  pages     = {231--248},
  year      = {2015},
  publisher = {Springer},
  address   = {Cham, Switzerland},
  doi       = {10.1007/978-3-319-22668-2_19},
  url       = {https://doi.org/10.1007/978-3-319-22668-2_19}
}

@book{schon1983reflective,
  author    = {Donald A. Sch{\"o}n},
  title     = {The Reflective Practitioner: How Professionals Think in Action},
  year      = {1983},
  publisher = {Basic Books},
  address   = {New York, NY, USA}
}

@article{gonzalez2025sense,
  title={Sense and Sensibility: What makes a social robot convincing to high-school students?},
  author={Gonz{\'a}lez-Oliveras, Pablo and Engwall, Olov and Majlesi, Ali Reza},
  journal={arXiv preprint arXiv:2506.12507},
  year={2025}
}

@article{risko2016cognitive,
  title={Cognitive offloading},
  author={Risko, Evan F and Gilbert, Sam J},
  journal={Trends in cognitive sciences},
  volume={20},
  number={9},
  pages={676--688},
  year={2016},
  publisher={Elsevier}
}

@article{bucinca2021trust,
  author    = {Zana Bu{\c{c}}inca and Maja Barbara Malaya and Krzysztof Z. Gajos},
  title     = {To Trust or to Think: Cognitive Forcing Functions Can Reduce
               Overreliance on {AI} in {AI}-Assisted Decision-Making},
  journal   = {Proceedings of the ACM on Human-Computer Interaction},
  volume    = {5},
  number    = {CSCW1},
  articleno = {188},
  pages     = {1--21},
  year      = {2021},
  publisher = {Association for Computing Machinery},
  address   = {New York, NY, USA},
  doi       = {10.1145/3449287},
  url       = {https://doi.org/10.1145/3449287}
}

@article{molenaar2022hybrid,
  author    = {Inge Molenaar},
  title     = {The Concept of Hybrid Human-{AI} Regulation:
               Exemplifying How to Support Young Learners'
               Self-Regulated Learning},
  journal   = {Computers and Education: Artificial Intelligence},
  volume    = {3},
  pages     = {100070},
  year      = {2022},
  publisher = {Elsevier},
  doi       = {10.1016/j.caeai.2022.100070},
  url       = {https://doi.org/10.1016/j.caeai.2022.100070}
}

@article{braun2019reflecting,
  author    = {Virginia Braun and Victoria Clarke},
  title     = {Reflecting on Reflexive Thematic Analysis},
  journal   = {Qualitative Research in Sport, Exercise and Health},
  volume    = {11},
  number    = {4},
  pages     = {589--597},
  year      = {2019},
  publisher = {Taylor \& Francis},
  doi       = {10.1080/2159676X.2019.1628806},
  url       = {https://doi.org/10.1080/2159676X.2019.1628806}
}

@article{pintrich2004conceptual,
  author    = {Paul R. Pintrich},
  title     = {A Conceptual Framework for Assessing Motivation and
               Self-Regulated Learning in College Students},
  journal   = {Educational Psychology Review},
  volume    = {16},
  number    = {4},
  pages     = {385--407},
  year      = {2004},
  publisher = {Springer},
  doi       = {10.1007/s10648-004-0006-x},
  url       = {https://doi.org/10.1007/s10648-004-0006-x}
}

@article{fan2025metacognitive,
  author    = {Yizhou Fan and Luzhen Tang and Huixiao Le and Kejie Shen
               and Shufang Tan and Yueying Zhao and Yuan Shen
               and Xinyu Li and Dragan Ga{\v{s}}evi{\'c}},
  title     = {Beware of Metacognitive Laziness: Effects of Generative
               Artificial Intelligence on Learning Motivation, Processes,
               and Performance},
  journal   = {British Journal of Educational Technology},
  volume    = {56},
  number    = {2},
  pages     = {489--530},
  year      = {2025},
  publisher = {Wiley},
  doi       = {10.1111/bjet.13544},
  url       = {https://doi.org/10.1111/bjet.13544}
}

@article{salah2026regulation,
  author    = {Mohammed Salah and Rudaina Al-Busaidi
               and Amna Al-Ghawi and Walid Shawky},
  title     = {The Regulation Paradox: Agentic {AI}, Bounded Autonomy,
               and Self-Regulated Learning in Higher Education},
  journal   = {Frontiers in Education},
  volume    = {11},
  pages     = {1902833},
  year      = {2026},
  publisher = {Frontiers Media SA},
  doi       = {10.3389/feduc.2026.1902833},
  url       = {https://doi.org/10.3389/feduc.2026.1902833}
}

@article{tsai2020empowering,
  author    = {Yi-Shan Tsai and Carlo Perrotta and Dragan Ga{\v{s}}evi{\'c}},
  title     = {Empowering Learners with Personalised Learning Approaches?
               Agency, Equity and Transparency in the Context of
               Learning Analytics},
  journal   = {Assessment \& Evaluation in Higher Education},
  volume    = {45},
  number    = {4},
  pages     = {554--567},
  year      = {2020},
  publisher = {Taylor \& Francis},
  doi       = {10.1080/02602938.2019.1676396},
  url       = {https://doi.org/10.1080/02602938.2019.1676396}
}

@article{lee2024harnessing,
  author    = {Sophia Soomin Lee and Robert L. Moore},
  title     = {Harnessing Generative {AI} ({GenAI}) for Automated Feedback
               in Higher Education: A Systematic Review},
  journal   = {Online Learning},
  volume    = {28},
  number    = {3},
  pages     = {82--106},
  year      = {2024},
  doi       = {10.24059/olj.v28i3.4593},
  url       = {https://doi.org/10.24059/olj.v28i3.4593}
}

@article{ryan2000self,
  author    = {Richard M. Ryan and Edward L. Deci},
  title     = {Self-Determination Theory and the Facilitation of Intrinsic
               Motivation, Social Development, and Well-Being},
  journal   = {American Psychologist},
  volume    = {55},
  number    = {1},
  pages     = {68--78},
  year      = {2000},
  publisher = {American Psychological Association},
  doi       = {10.1037/0003-066X.55.1.68},
  url       = {https://doi.org/10.1037/0003-066X.55.1.68}
}

@book{vygotsky1978mind,
  author    = {Lev S. Vygotsky},
  title     = {Mind in Society: The Development of Higher Psychological Processes},
  editor    = {Michael Cole and Vera John-Steiner and Sylvia Scribner and Ellen Souberman},
  year      = {1978},
  publisher = {Harvard University Press},
  address   = {Cambridge, MA, USA},
  isbn      = {9780674576292}
}

@article{braun2006thematic,
  author    = {Virginia Braun and Victoria Clarke},
  title     = {Using Thematic Analysis in Psychology},
  journal   = {Qualitative Research in Psychology},
  volume    = {3},
  number    = {2},
  pages     = {77--101},
  year      = {2006},
  publisher = {Taylor \& Francis},
  doi       = {10.1191/1478088706qp063oa},
  url       = {https://doi.org/10.1191/1478088706qp063oa}
}

@article{wegerif2026dialogic,
  author    = {Rupert Wegerif},
  title     = {Dialogic Intelligence: Rethinking What Education Is for
               in the Age of {AI} / Inteligencia Dial{\'o}gica:
               Repensando el Prop{\'o}sito de la Educaci{\'o}n
               en la Era de la {IA}},
  journal   = {Journal for the Study of Education and Development},
  volume    = {49},
  number    = {3},
  pages     = {616--643},
  year      = {2026},
  publisher = {SAGE Publications},
  doi       = {10.1177/02103702261439739},
  url       = {https://doi.org/10.1177/02103702261439739}
}

@inproceedings{park2026llmbox,
  author    = {Sohyun Park and Tak Yeon Lee and Woohun Lee},
  title     = {{LLM}-box vs. Thinking-box: Designing for Deliberate
               User Engagement with Distorted Information
               in Conversational Search},
  booktitle = {Proceedings of the 2026 CHI Conference on Human Factors
               in Computing Systems},
  series    = {CHI '26},
  year      = {2026},
  articleno = {367},
  numpages  = {20},
  publisher = {Association for Computing Machinery},
  address   = {New York, NY, USA},
  doi       = {10.1145/3772318.3790271},
  url       = {https://doi.org/10.1145/3772318.3790271}
}

@inproceedings{gmeiner2025metacognitive,
  author    = {Frederic Gmeiner and Kaitao Luo and Ye Wang
               and Kenneth Holstein and Nikolas Martelaro},
  title     = {Exploring the Potential of Metacognitive Support Agents
               for Human-{AI} Co-Creation},
  booktitle = {Proceedings of the 2025 ACM Designing Interactive Systems
               Conference},
  series    = {DIS '25},
  year      = {2025},
  pages     = {1244--1269},
  publisher = {Association for Computing Machinery},
  address   = {New York, NY, USA},
  doi       = {10.1145/3715336.3735785},
  url       = {https://doi.org/10.1145/3715336.3735785}
}

@inproceedings{cheng2025noticeability,
  author    = {Yi Fei Cheng and David Lindlbauer},
  title     = {Sensing Noticeability in Ambient Information Environments},
  booktitle = {Proceedings of the 2025 CHI Conference on Human Factors
               in Computing Systems},
  series    = {CHI '25},
  year      = {2025},
  articleno = {810},
  numpages  = {17},
  publisher = {Association for Computing Machinery},
  address   = {New York, NY, USA},
  doi       = {10.1145/3706598.3713511},
  url       = {https://doi.org/10.1145/3706598.3713511}
}

@article{darvishi2024agency,
  author    = {Ali Darvishi and Hassan Khosravi and Shazia Sadiq
               and Dragan Ga{\v{s}}evi{\'c} and George Siemens},
  title     = {Impact of {AI} Assistance on Student Agency},
  journal   = {Computers \& Education},
  volume    = {210},
  pages     = {104967},
  year      = {2024},
  publisher = {Elsevier},
  doi       = {10.1016/j.compedu.2023.104967},
  url       = {https://doi.org/10.1016/j.compedu.2023.104967}
}

@misc{weiser1996calm,
  author       = {Mark Weiser and John Seely Brown},
  title        = {The Coming Age of Calm Technology},
  year         = {1996},
  month        = oct,
  howpublished = {Xerox PARC},
  note         = {October 5, 1996},
  url          = {https://calmtech.com/papers/coming-age-calm-technology}
}

@article{borner2014lead,
  author    = {Dirk B{\"o}rner and Marco Kalz and Marcus Specht},
  title     = {Lead Me Gently: Facilitating Knowledge Gain through
               Attention-Aware Ambient Learning Displays},
  journal   = {Computers \& Education},
  volume    = {78},
  pages     = {10--19},
  year      = {2014},
  publisher = {Elsevier},
  doi       = {10.1016/j.compedu.2014.04.017},
  url       = {https://doi.org/10.1016/j.compedu.2014.04.017}
}

@inproceedings{liu2026tasklens,
  author    = {Yimeng Liu and Misha Sra},
  title     = {{TaskLens}: Generating Task-Conditioned Scaffolded Interfaces
               for Learning Professional Creative Software},
  booktitle = {Proceedings of the 2026 Designing Interactive Systems Conference},
  series    = {DIS '26},
  year      = {2026},
  pages     = {17--38},
  publisher = {Association for Computing Machinery},
  address   = {New York, NY, USA},
  doi       = {10.1145/3800645.3813081},
  url       = {https://doi.org/10.1145/3800645.3813081}
}

@inproceedings{zhang2025friction,
  author    = {Chao Zhang and Kexin Ju and Peter Bidoshi
               and Yu-Chun Grace Yen and Jeffrey M. Rzeszotarski},
  title     = {Friction: Deciphering Writing Feedback into Writing Revisions
               through {LLM}-Assisted Reflection},
  booktitle = {Proceedings of the 2025 CHI Conference on Human Factors
               in Computing Systems},
  series    = {CHI '25},
  year      = {2025},
  articleno = {935},
  numpages  = {27},
  publisher = {Association for Computing Machinery},
  address   = {New York, NY, USA},
  doi       = {10.1145/3706598.3714316},
  url       = {https://doi.org/10.1145/3706598.3714316}
}

@inproceedings{anik2026dataset,
  author    = {Ariful Islam Anik and Andrea Bunt},
  title     = {Designing Effective Training Dataset Explanations:
               The Impact of Information Depth and Progressive Disclosure},
  booktitle = {Proceedings of the 31st International Conference on
               Intelligent User Interfaces},
  series    = {IUI '26},
  year      = {2026},
  pages     = {426--441},
  publisher = {Association for Computing Machinery},
  address   = {New York, NY, USA},
  doi       = {10.1145/3742413.3789087},
  url       = {https://doi.org/10.1145/3742413.3789087}
}

@article{zhao2024gamifiedvr,
  author    = {Jia-Hua Zhao and Patcharin Panjaburee
               and Gwo-Jen Hwang and Wararat Wongkia},
  title     = {Effects of a Self-Regulated-Based Gamified Virtual Reality
               System on Students' English Learning Performance and Affection},
  journal   = {Interactive Learning Environments},
  volume    = {32},
  number    = {9},
  pages     = {5509--5536},
  year      = {2024},
  publisher = {Routledge},
  doi       = {10.1080/10494820.2023.2219702},
  url       = {https://doi.org/10.1080/10494820.2023.2219702}
}

@article{zhang2026its,
  author    = {Kaiyu Zhang and Ying Zhou and Cheng Zhang and Jihe Chen},
  title     = {The Effects of Intelligent Tutoring Systems on
               Self-Regulated Learning: A Meta-Analysis},
  journal   = {Educational Psychology Review},
  volume    = {38},
  number    = {1},
  year      = {2026},
  publisher = {Springer},
  doi       = {10.1007/s10648-026-10199-4},
  url       = {https://doi.org/10.1007/s10648-026-10199-4}
}

@article{wang2025srlmentor,
  author    = {Sikai Wang and Xinyi Luo and Khe Foon Hew},
  title     = {Enhancing Self-Regulated Learning Using Generative {AI}:
               Development and Evaluation of {SRLMentor}},
  journal   = {IEEE Transactions on Learning Technologies},
  volume    = {18},
  pages     = {1048--1061},
  year      = {2025},
  publisher = {IEEE},
  doi       = {10.1109/TLT.2025.3634216},
  url       = {https://doi.org/10.1109/TLT.2025.3634216}
}

@article{wu2026interactive,
  author    = {Henan Wu and Zilin Wang and Xinyuan Wen},
  title     = {Examining Interactive {AI}-Supported Learning Environments:
               The Mediating Role of {AI} Metacognition and the Moderating
               Role of Cognitive Flexibility},
  journal   = {Acta Psychologica},
  volume    = {266},
  pages     = {106874},
  year      = {2026},
  publisher = {Elsevier},
  doi       = {10.1016/j.actpsy.2026.106874},
  url       = {https://doi.org/10.1016/j.actpsy.2026.106874}
}

@techreport{taylor2026contextmaxxing,
  author      = {Jacob Taylor and Kershlin Krishna},
  title       = {Context-maxxing: A Path to Cognitive Agency with
                 Generative {AI}},
  institution = {Brookings Institution},
  type        = {Working Paper},
  year        = {2026},
  month       = may,
  url         = {https://www.brookings.edu/articles/context-maxxing-cognitive-agency-generative-ai/}
}

@inproceedings{chucarroll1997tracking,
  author    = {Jennifer Chu-Carroll and Michael K. Brown},
  title     = {Tracking Initiative in Collaborative Dialogue Interactions},
  booktitle = {35th Annual Meeting of the Association for Computational Linguistics
               and 8th Conference of the European Chapter of the Association
               for Computational Linguistics},
  year      = {1997},
  pages     = {262--270},
  publisher = {Association for Computational Linguistics},
  address   = {Madrid, Spain},
  doi       = {10.3115/976909.979651},
  url       = {https://doi.org/10.3115/976909.979651}
}

@article{hayes1986writing,
  author    = {John R. Hayes and Linda S. Flower},
  title     = {Writing Research and the Writer},
  journal   = {American Psychologist},
  volume    = {41},
  number    = {10},
  pages     = {1106--1113},
  year      = {1986},
  publisher = {American Psychological Association},
  doi       = {10.1037/0003-066X.41.10.1106},
  url       = {https://doi.org/10.1037/0003-066X.41.10.1106}
}

@inproceedings{isa2026codegeneration,
  author    = {Hajara-Yasmin Isa and Matthew Weston and Muhammad Rizky Wellyanto
               and Ishita Karna and Jerry O. Talton and Ranjitha Kumar},
  title     = {From Code Generation to Conceptual Learning:
               Student Use of {LLM}s in a Web Programming Course},
  booktitle = {Proceedings of the 2026 CHI Conference on Human Factors
               in Computing Systems},
  series    = {CHI '26},
  year      = {2026},
  articleno = {1598},
  numpages  = {13},
  publisher = {Association for Computing Machinery},
  address   = {New York, NY, USA},
  doi       = {10.1145/3772318.3793207},
  url       = {https://doi.org/10.1145/3772318.3793207}
}

@inproceedings{jelson2026empirical,
  author    = {Andrew Jelson and Daniel Manesh and Alice Jang
               and Daniel Dunlap and Young-Ho Kim and Sang Won Lee},
  title     = {An Empirical Study to Understand How Students Use
               {ChatGPT} for Writing Essays},
  booktitle = {Proceedings of the 2026 CHI Conference on Human Factors
               in Computing Systems},
  series    = {CHI '26},
  year      = {2026},
  publisher = {Association for Computing Machinery},
  address   = {New York, NY, USA},
  doi       = {10.1145/3772318.3791056},
  url       = {https://doi.org/10.1145/3772318.3791056}
}

@inproceedings{yun2026chatbot,
  author    = {Bhada Yun and Evgenia Taranova and April Yi Wang},
  title     = {Does My Chatbot Have an Agenda? Understanding Human and
               {AI} Agency in Human-Human-like Chatbot Interaction},
  booktitle = {Proceedings of the 2026 CHI Conference on Human Factors
               in Computing Systems},
  series    = {CHI '26},
  year      = {2026},
  articleno = {373},
  numpages  = {32},
  publisher = {Association for Computing Machinery},
  address   = {New York, NY, USA},
  doi       = {10.1145/3772318.3791620},
  url       = {https://doi.org/10.1145/3772318.3791620}
}

@inproceedings{wu2026afterinterface,
  author    = {Mengke Wu and Mike Yao},
  title     = {After the Interface: Relocating Human Agency in the Age of
               Conversational {AI}},
  booktitle = {Proceedings of the 8th ACM Conference on Conversational
               User Interfaces},
  series    = {CUI '26},
  year      = {2026},
  pages     = {1--7},
  publisher = {Association for Computing Machinery},
  address   = {New York, NY, USA},
  doi       = {10.1145/3816046.3816301},
  url       = {https://doi.org/10.1145/3816046.3816301}
}

@inproceedings{pauzi2026evaluating,
  author    = {Zaki Pauzi and Vidminas Vizgirda
               and Leslye Denisse Dias Duran and Manolis Mavrikis},
  title     = {Towards Evaluating Student Agency in {AI}-Mediated Learning},
  booktitle = {Companion Publication of the 2026 ACM Designing
               Interactive Systems Conference},
  series    = {DIS '26 Companion},
  year      = {2026},
  pages     = {221--225},
  publisher = {Association for Computing Machinery},
  address   = {New York, NY, USA},
  doi       = {10.1145/3802974.3809419},
  url       = {https://doi.org/10.1145/3802974.3809419}
}

@inproceedings{bennett2023agency,
  author    = {Dan Bennett and Oussama Metatla and Anne Roudaut and Elisa D. Mekler},
  title     = {How does {HCI} Understand Human Agency and Autonomy?},
  booktitle = {Proceedings of the 2023 CHI Conference on Human Factors
               in Computing Systems},
  series    = {CHI '23},
  year      = {2023},
  pages     = {1--18},
  publisher = {Association for Computing Machinery},
  address   = {New York, NY, USA},
  doi       = {10.1145/3544548.3580651},
  url       = {https://doi.org/10.1145/3544548.3580651}
}

@article{zhang2025collaboration,
  author    = {Shuning Zhang and Hui Wang and Xin Yi},
  title     = {Exploring Collaboration Patterns and Strategies in Human-{AI}
               Co-creation through the Lens of Agency: A Scoping Review of
               the Top-tier {HCI} Literature},
  journal   = {Proceedings of the ACM on Human-Computer Interaction},
  volume    = {9},
  number    = {7},
  year      = {2025},
  articleno = {CSCW413},
  numpages  = {43},
  publisher = {Association for Computing Machinery},
  address   = {New York, NY, USA},
  doi       = {10.1145/3757594},
  url       = {https://doi.org/10.1145/3757594}
}

@inproceedings{song2026goodfences,
  author    = {Jiwon Song and Aeri Cho and Sihyeon Lee and Kiroong Choe and Jinwook Seo},
  title     = {Good Fences Make Good Learning: How Self-Directed Language Learners
               Navigate {LLM} Delegation Decisions},
  booktitle = {Proceedings of the 2026 CHI Conference on Human Factors
               in Computing Systems},
  series    = {CHI '26},
  year      = {2026},
  articleno = {82},
  numpages  = {18},
  publisher = {Association for Computing Machinery},
  address   = {New York, NY, USA},
  doi       = {10.1145/3772318.3791657},
  url       = {https://doi.org/10.1145/3772318.3791657}
}

@inproceedings{borchers2026whodecides,
  author    = {Conrad Borchers and Olga Viberg and Ren{\'e} F. Kizilcec},
  title     = {Who Decides in {AI}-Mediated Learning? The Agency Allocation Framework},
  booktitle = {Proceedings of the Thirteenth ACM Conference on Learning @ Scale},
  series    = {L@S '26},
  year      = {2026},
  pages     = {22--33},
  publisher = {Association for Computing Machinery},
  address   = {New York, NY, USA},
  doi       = {10.1145/3774398.3811627},
  url       = {https://doi.org/10.1145/3774398.3811627}
}

@inproceedings{sharma2026hybrid,
  author    = {Paras Sharma and Yueping Sha and Janet Shufor Bih Epse Fofang
               and Brayden Yan and Jess A. Turner and Nicole Balay
               and Hubert O. Asare and Angela E. B. Stewart and Erin Walker},
  title     = {Hybrid {LLM}-Embedded Dialogue Agents for Learner Reflection:
               Designing Responsive and Theory-Driven Interactions},
  booktitle = {Proceedings of the 2026 CHI Conference on Human Factors
               in Computing Systems},
  series    = {CHI '26},
  year      = {2026},
  articleno = {137},
  numpages  = {20},
  publisher = {Association for Computing Machinery},
  address   = {New York, NY, USA},
  doi       = {10.1145/3772318.3791582},
  url       = {https://doi.org/10.1145/3772318.3791582}
}

@inproceedings{li2026supporting,
  author    = {Tiffany Wenting Li and Yifan Song and Hari Sundaram
               and Karrie Karahalios},
  title     = {Supporting Learners' Use of Imperfect Generative Pedagogical Chatbots: The Role of Chatbot Response Uncertainty and Reduced Verbosity},
  booktitle = {Proceedings of the 2026 CHI Conference on Human Factors
               in Computing Systems},
  series    = {CHI '26},
  year      = {2026},
  articleno = {704},
  publisher = {Association for Computing Machinery},
  address   = {New York, NY, USA},
  doi       = {10.1145/3772318.3791940},
  url       = {https://doi.org/10.1145/3772318.3791940}
}

@inproceedings{klein2026answer,
  author    = {Reinhard Klein and Daria Benden and Alexander Schier
               and David Stotko and Fani Lauermann},
  title     = {From Answer Engines to Learning Partners: A Dual-{ZPD}
               Design Framework for {AI}-Supported Learning},
  booktitle = {Proceedings of the 2026 CHI Conference on Human Factors
               in Computing Systems},
  series    = {CHI '26},
  year      = {2026},
  numpages  = {16},
  publisher = {Association for Computing Machinery},
  address   = {New York, NY, USA},
  doi       = {10.1145/3772318.3791476},
  url       = {https://doi.org/10.1145/3772318.3791476}
}

@inproceedings{chen2026passive,
  author    = {Zixin Chen and Haotian Li and Zhe Liu and Huamin Qu
               and Xing Xie},
  title     = {From Passive Consumption to Active Interaction: Exploring
               Interactive {LLM} Scaffolding to Support Learning Engagement},
  booktitle = {Extended Abstracts of the 2026 CHI Conference on Human
               Factors in Computing Systems},
  series    = {CHI EA '26},
  year      = {2026},
  articleno = {324},
  numpages  = {6},
  publisher = {Association for Computing Machinery},
  address   = {New York, NY, USA},
  doi       = {10.1145/3772363.3798931},
  url       = {https://doi.org/10.1145/3772363.3798931}
}

@inproceedings{tian2026critical,
  author    = {Harry Yizhou Tian and Hasan Amin and Ming Yin},
  title     = {Understanding the Effects of {AI}-Assisted Critical Thinking
               on Human-{AI} Decision Making},
  booktitle = {Proceedings of the 2026 CHI Conference on Human Factors
               in Computing Systems},
  series    = {CHI '26},
  year      = {2026},
  articleno = {810},
  numpages  = {30},
  publisher = {Association for Computing Machinery},
  address   = {New York, NY, USA},
  doi       = {10.1145/3772318.3790785},
  url       = {https://doi.org/10.1145/3772318.3790785}
}

@inproceedings{dhillon2024shaping,
  author    = {Paramveer S. Dhillon and Somayeh Molaei and Jiaqi Li
               and Maximilian Golub and Shaochun Zheng
               and Lionel P. Robert Jr.},
  title     = {Shaping Human-{AI} Collaboration: Varied Scaffolding Levels
               in Co-writing with Language Models},
  booktitle = {Proceedings of the 2024 CHI Conference on Human Factors
               in Computing Systems},
  series    = {CHI '24},
  year      = {2024},
  articleno = {1044},
  numpages  = {18},
  publisher = {Association for Computing Machinery},
  address   = {New York, NY, USA},
  doi       = {10.1145/3613904.3642134},
  url       = {https://doi.org/10.1145/3613904.3642134}
}

@inproceedings{viberg2026scalable,
  author    = {Olga Viberg and Jacqueline Wong and Selma Ozdere
               and Richard Lee Davis},
  title     = {Scalable {SRL} Conversational Scaffolding for
               Student--{LLM} Interaction},
  booktitle = {Proceedings of the Thirteenth ACM Conference on
               Learning at Scale},
  series    = {L@S '26},
  year      = {2026},
  pages     = {545--549},
  publisher = {Association for Computing Machinery},
  address   = {New York, NY, USA},
  doi       = {10.1145/3774398.3811582},
  url       = {https://doi.org/10.1145/3774398.3811582}
}

@inproceedings{fan2026learneragency,
  author    = {Yueling Fan and Jacqueline Wong and Olov Engwall and Olga Viberg},
  title     = {Towards Learner Agency-Preserving Feedback in Generative {AI}: A Participatory Design Study},
  booktitle = {Proceedings of the European Conference on Technology Enhanced Learning},
  year      = {2026},
  note = {In press}
}

@inproceedings{borchers2026agency,
  author    = {Conrad Borchers and Olga Viberg and Ren{\'e} F. Kizilcec},
  title     = {Who Decides in {AI}-Mediated Learning?
               The Agency Allocation Framework},
  booktitle = {Proceedings of the Thirteenth ACM Conference on
               Learning at Scale},
  series    = {L@S '26},
  year      = {2026},
  pages     = {22--33},
  publisher = {Association for Computing Machinery},
  address   = {New York, NY, USA},
  doi       = {10.1145/3774398.3811627},
  url       = {https://doi.org/10.1145/3774398.3811627}
}

\appendix

\end{document}